\documentclass[12pt,preprint,numberedappendix,twocolappendix]{emulateapj}

\newcommand\bibinc{n}		

\usepackage{lineno}

\usepackage{blindtext}
\usepackage{subeqnarray}
\usepackage{amsmath}
\usepackage{hyperref}
\usepackage{ulem}
\usepackage{stmaryrd}

\hypersetup{colorlinks=true,linkcolor=black,citecolor=blue,urlcolor=red}

\usepackage{natbib}
\def\tt{ \bf }

\newcommand{\trad}{\tau_{\rm{rad}}}

\def \teq {T_\mathrm{eq}}

\def \mmkg {\rm m^2kg^{-1}}
\def \btt {}
\def \bttt {}

\setcitestyle{citesep={,}}
\begin{document}


\shorttitle{Seasonality on exoplanets}
\shortauthors{X. Tan}

\title{Weak seasonality on temperate exoplanets around low-mass stars}
\author{Xianyu Tan}
\affil{Atmospheric, Oceanic  and Planetary Physics, Department of Physics, University of Oxford, OX1 3PU, UK\\
\url{xianyu.tan@physics.ox.ac.uk}}

\begin{abstract}
Planets with non-zero obliquity and/or orbital eccentricity experience  seasonal variations of stellar irradiation {\btt at local latitudes}. The extent of the atmospheric response can be crudely estimated by the ratio between the orbital timescale and the atmospheric radiative timescale. Given a set of atmospheric parameters, we show that this ratio  depends mostly on the stellar properties and is independent of orbital distance  and planetary equilibrium temperature. For Jupiter-like atmospheres, this ratio is $\ll1$ for planets around very-low-mass M dwarfs and $\gtrsim1$ when the stellar mass is greater than about 0.6 solar mass. Complications can arise from various factors, including varying atmospheric metallicity, clouds and atmospheric dynamics. {\btt Given the eccentricity and obliquity,} the seasonal response is expected to be systematically weaker for gaseous exoplanets around low-mass stars and stronger  for  those  around  more massive  stars. The amplitude  and phase lag of atmospheric seasonal variations as a function of host stellar mass  are quantified by   idealized analytic models. {\btt  At the infrared emission level in the photosphere}, the relative amplitudes  of thermal flux and temperature perturbations are negligible, and their phase lags are closed to $-90^{\circ}$ for Jupiter-like planets around very-low-mass stars. The relative amplitudes and phase lags increase gradually with increasing stellar mass. With a particular stellar mass, the relative amplitude and phase lag decrease  from low to high infrared optical depth. {\btt We also present numerical calculations for a better illustration of the seasonal behaviors.} Lastly, we discuss  implications for the atmospheric circulation   and  future atmospheric characterization of exoplanets in systems with different stellar masses.
\end{abstract}

\section{introduction}
{\btt Planetary seasonality generally arises from the  change of  stellar irradiation at certain latitudes over the orbital timescale in the precence of non-zero planetary obliquity  or orbital eccentricity. } Solar-system planets with non-zero obliquity  exhibit significant seasonal variations, {\btt which have been observed on Saturn (see reviews by \citealp{fletcher2018,fletcher2020}) and Uranus (see a review by \citealp{fletcher2021}). Even Jupiter, which has a negligible obliquity of $\sim 3^{\circ}$, shows evidence of an eccentricity season \citep{nixon2010}. The orbital configurations of exoplanets are more diverse, and we expect richer seasonal behaviors in the atmospheres of exoplanets. }

Previous work has studied  the climate and seasonal dynamics of exoplanets  and the implications for observation and habitability. A branch of these studies has focused on modeling the climate and atmospheric dynamics of terrestrial planets with Earth-like atmospheres (e.g., \citealp{williams1997, williams2002, gaidos2004, ferreira2014,shields2016, guendelman2019,kang2019,kang2019b, palubski2020}). The second category contains investigations of variability,  atmospheric general circulation and potential observable signatures of giant planets  with non-zero obliquity  \citep{langton2007,rauscher2017, ohno2019,ohno2019b} and high eccentricities \citep{langton2008,lewis2010, kataria2013,lewis2014, mayorga2021}.  Some others do not directly address the climate and seasonal dynamics but try to identify means by which we could measure or constrain planetary obliquity of exoplanets (e.g., \citealp{kawahara2010,cowan2013,schwartz2016,luger2021}).

Studies of the seasonality of exoplanets have mostly focused on specific targets  or systems around certain types of stars. In this work, we show that, {\btt although the  seasonal variation of incoming stellar flux is mainly controlled by the local latitude, obliquity and eccentricity, given a type of atmosphere, the characteristic atmospheric response (which may be represented as nondimensional quantities  related to the variation of stellar flux) to such seasonal variations  } primarily depends on the stellar properties and is independent of the orbital distance  and planetary equilibrium temperature.  Variations of the mean thermal opacity and specific heat introduce anomalies of the seasonal response,  but   the seasonal response is expected to be systematically  weaker on exoplanets around low-mass stars and stronger on those around higher-mass stars.  

Throughout this study, we demonstrate this point for planets with   moderate equilibrium temperatures and thick hydrogen-dominated atmospheres, partly because the gaseous planets are more likely to be characterized in the near future via upcoming missions such as the Nancy Grace Roman Space Telescope \citep{spergel2013wide} and mission concepts such as LUVOIR \citep{bolcar2016initial}  or HabEx \citep{mennesson2016habitable} that aim to measure reflected lights of cold and temperate exoplanets by direct imaging. {\btt We expect that a large fraction of temperate gaseous planets should have sufficient orbital distances to avoid significant obliquity tidal damping by the host stars  if they acquired non-zero obliquity during their formation and evolution.      Our analysis does not apply to close-in planets that may have been tidally locked\footnote{\btt Some close-in planets may still   have chances to maintain non-zero obliquity through perturbations from additional companions in the system \citep{su2021}. }.  } Atmospheres of terrestrial planets likely have more diversities, including varying atmospheric mass, composition and surface conditions, which sufficiently complicate the simple trend that holds  for thick hydrogen-dominated atmospheres. 

This paper is organized as follows. In Section \ref{ch.timescale}, we carry out the essential argument using comparisons between the orbital and atmospheric radiative timescales. In Section \ref{ch.rt}, we further present  idealized  analytic and numerical radiative transfer calculations to support  results in Section \ref{ch.timescale} and also show certain quantitative differences. Finally, we discuss limitations  and implications of our results in Section \ref{ch.discussion}.

\section{Timescale comparison}
\label{ch.timescale}
{\btt The characteristic atmospheric response to the seasonally varying irradiation at certain latitudes can be intuitively understood by the comparison between the orbital timescale and the radiative timescale \citep{ohno2019},} the latter of which is  the characteristic  timescale it would take for an atmosphere to relax its temperature back to the radiative equilibrium via thermal radiation. The amplitude of the response is expected to be smaller  when the radiative timescale is much longer than the orbital timescale, and vice versa.

With the cool-to-space approximation (e.g., \citealp{andrews1987,showman2002}), the radiative timescale is $\trad\sim\frac{p}{g}\frac{c_p}{4\sigma \teq^3}$, {\btt where $p$ is pressure where the infrared optical depth is around unity}, $g$ is surface gravity, $c_p$ is specific heat at constant pressure, and $\sigma$ is the Stefan-Boltzmann constant. {\btt This formula ceases to be valid at pressures that significantly deviate from the thermal photospheric pressure \citep{showman2002}.} Near the mean infrared (IR) photosphere where the mean IR optical depth is 1, the radiative timescale is then $\trad\sim\frac{c_p}{4\kappa\sigma \teq^3}$, where $\kappa$ is the mean IR opacity.  Here, $\teq$ is the planetary equilibrium temperature defined as a homogeneous blackbody temperature  the planet would have in order to globally radiate away the  annual-mean stellar irradiation:
\begin{equation}
\begin{split}
    4\pi r^2_p \sigma \teq^4 = \langle F_{\star}\rangle, \\
    \langle F_{\star}\rangle = \frac{1-A}{\sqrt{1-e^2}}\frac{\sigma T^4_{\star}R^2_{\star}}{a^2}\pi r^2_p,
    \end{split}
    \label{eq.teq}
\end{equation}
where $r_p$ is the planetary radius, $a$ is the semimajor axis, $e$ is the eccentricity, $A$ is the planetary bond albedo, $T_{\star}$ is the stellar effective temperature, and $R_{\star}$ is the stellar radius. The annual-mean stellar flux $\langle F_{\star}\rangle$ with non-zero eccentricity has been well known (e.g., \citealp{williams2002}).
Using Kepler's third law $P_{\rm orbit}^2\approx\frac{4\pi^2}{GM_{\star}}a^3$ where $P_{\rm orbit}$ is the orbital period, $M_{\star}$ is the stellar mass\footnote{\btt When the planetary mass $M_{\rm p}$ is a nontrivial fraction of the total mass of the system, one should use $P_{\rm orbit}^2=\frac{4\pi^2}{G(M_{\star}+M_{\rm p})}a^3$.} and $G$ is the gravitational constant,  the ratio between the orbital period and the radiative timescale is
\begin{equation}
\begin{split}
    \frac{P_{\rm orbit}}{\trad}   = &~ \frac{\sqrt{8}\pi\sigma\kappa T_{\star}^3R_{\star}^{3/2}}{c_p\sqrt{GM_{\star}}}\left(\frac{1-A}{\sqrt{1-e^2}}\right)^{\frac{3}{4}} \\
    \approx &~ 5.94 \left(\frac{\kappa}{5\times10^{-4}}\right) \left(\frac{c_p}{13000}\right)^{-1}\left(\frac{T_{\star}}{T_{\odot}}\right)^3  \\
    & \times \left(\frac{R_{\star}}{R_{\odot}}\right)^{\frac{3}{2}} \left(\frac{M_{\star}}{M_{\odot}}\right)^{-\frac{1}{2}}\left(\frac{1-A}{\sqrt{1-e^2}}\right)^{\frac{3}{4}},
    \end{split}
    \label{eq.essence}
\end{equation}
where $\kappa$ and $c_p$ are in SI unit, $T_{\odot}$, $R_{\odot}$ and $M_{\odot}$ are  the effective temperature, radius and mass of the sun  which are adopted as 5772 K, $6.957\times10^8$ m and $1.9891 \times 10^{30}$ kg, respectively.
 
For a particular set of atmospheric parameters of $\kappa$, $c_p$ and $A$,  $P_{\rm orbit}/\trad$ increases rapidly  with increasing stellar luminosity and decreases with stellar mass. This ratio is primarily determined by properties of the host star (i.e., stellar mass, effective temperature and radius) and also by the eccentricity to some extent. This suggests that the stellar host is essential for the planetary seasonality for all  planets that have a certain type of atmosphere in the system  regardless their orbital distance  and effective temperature. Below we discuss certain complications. The mean thermal opacity increases with increasing metallicity in the atmosphere (e.g., \citealp{freedman2014}), which could span more than one order of magnitude in atmospheres of gaseous planets (e.g., \citealp{kreidberg2014,line2021}). Extra complications  come  from condensation. For instance, water condenses out in the mean thermal photospheres of all solar-system gaseous planets. This  reduces significant thermal opacity  from water vapor, but at the same time loads thermal and visible opacity   from   cloud particles  which can be highly uncertain and inhomogeneous. The change of specific heat is not expected to  be significant as long as the atmosphere is dominated by hydrogen and helium. All major planetary atmospheres in the solar system exhibit non-zero bond albedo, and  their major sources  include Rayleigh scattering and scattering due to clouds and aerosoles \citep{conrath1989}. For exoplanets, the bond albedo is  highly uncertain and  may vary from negligible to a fraction of one (e.g., \citealp{schwartz2015}).  In contrast to solar-system planets, the eccentricities of exoplanets can span a wide range (see http://exoplanets.org/) which could increase  the variation of $P_{\rm orbit}/\trad$.  

\begin{figure*}      
\epsscale{1.15 }      
\centering
\plotone{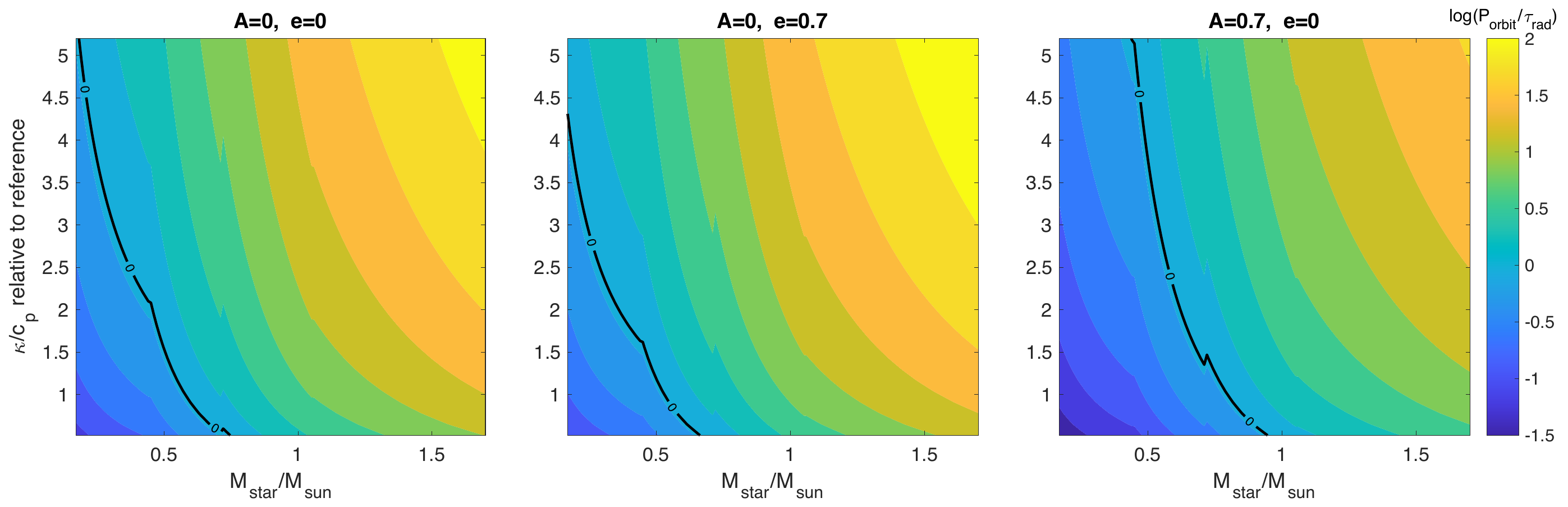}
\caption{  The logarithm of the ratio between the orbital period and the radiative timescale  at $\tau=1$ level, $\log(P_{\rm orbit}/\trad)$, as a function of stellar mass and  $\kappa/c_p$ relative to the reference value (the reference infrared opacity is $\kappa_{\rm ref}=5\times10^{-4}\;{\rm m^2kg^{-1}}$ and specific heat is $c_{p,{\rm ref}}=  13000 \;{\rm Jkg^{-1}K^{-1}}$). The reference values are appropriate for Jovian atmosphere. Different bond albedo $A$ and eccentricity $e$ are assumed for results in different panels as labeled above. Stellar radius and effective temperature as a function of stellar mass are adopted from \cite{eker2018}. Black lines represent contours of  $P_{\rm orbit}/\trad=1$. Discontinuity in the plots arises from the segmental fits in \cite{eker2018}. 
}
\label{fig.compare2D}
\end{figure*}

Given the complications mentioned above,   we compute  $P_{\rm orbit}/\trad$ as a function of varying $\kappa/c_p$ and stellar mass.     Results are  shown in Figure \ref{fig.compare2D} as a function of stellar mass and $\kappa/c_p$ relative to a reference value, and each panel contains results with a set of  eccentricity $e$ and bond albedo $A$.  We adopt Jupiter-like reference values of  $\kappa_{\rm ref}=5\times10^{-4}\;\mmkg$ and $c_{p, {\rm ref}}=13000 \;{\rm Jkg^{-1}K^{-1}}$. This $\kappa_{\rm ref}$ results in a mean thermal photosphere at about 0.5 bar for Jovian gravity which is reasonable for Jovian-like conditions \citep{Sromovsky1998}. The relations between stellar radius, effective temperature and stellar mass are adopted from the fits in \cite{eker2018} that are based on a large sample of stars in the solar neighbourhood.\footnote{Since this is an order-of-magnitude estimation, a slight change of database of stellar properties is not expected to change our qualitative results and conclusions.} At a given $\kappa/c_p$,  $P_{\rm orbit}/\trad$  spans about two orders of magnitude when the stellar mass varies from 0.18 to 1.7 $M_{\odot}$. For $\kappa/c_p$ near the reference value,  $P_{\rm orbit}/\trad$  is much less than 1 for low-mass M dwarfs and $> 1$ with stellar masses $\gtrapprox 0.6\;M_{\odot}$ with moderate bond albedo and eccentricity.  At sufficiently high  $\kappa/c_p$ which represents high metallicity in the atmosphere, $P_{\rm orbit}/\trad$ is generally close to or larger than 1, suggesting that metal-rich atmospheres tend to hold strong seasonality regardless the stellar mass. Nevertheless,   seasonality  on  gaseous planets around low-mass stars should be statistically weaker  than those  orbiting more massive stars. 

\begin{figure}      
\epsscale{1.}      
\centering
\plotone{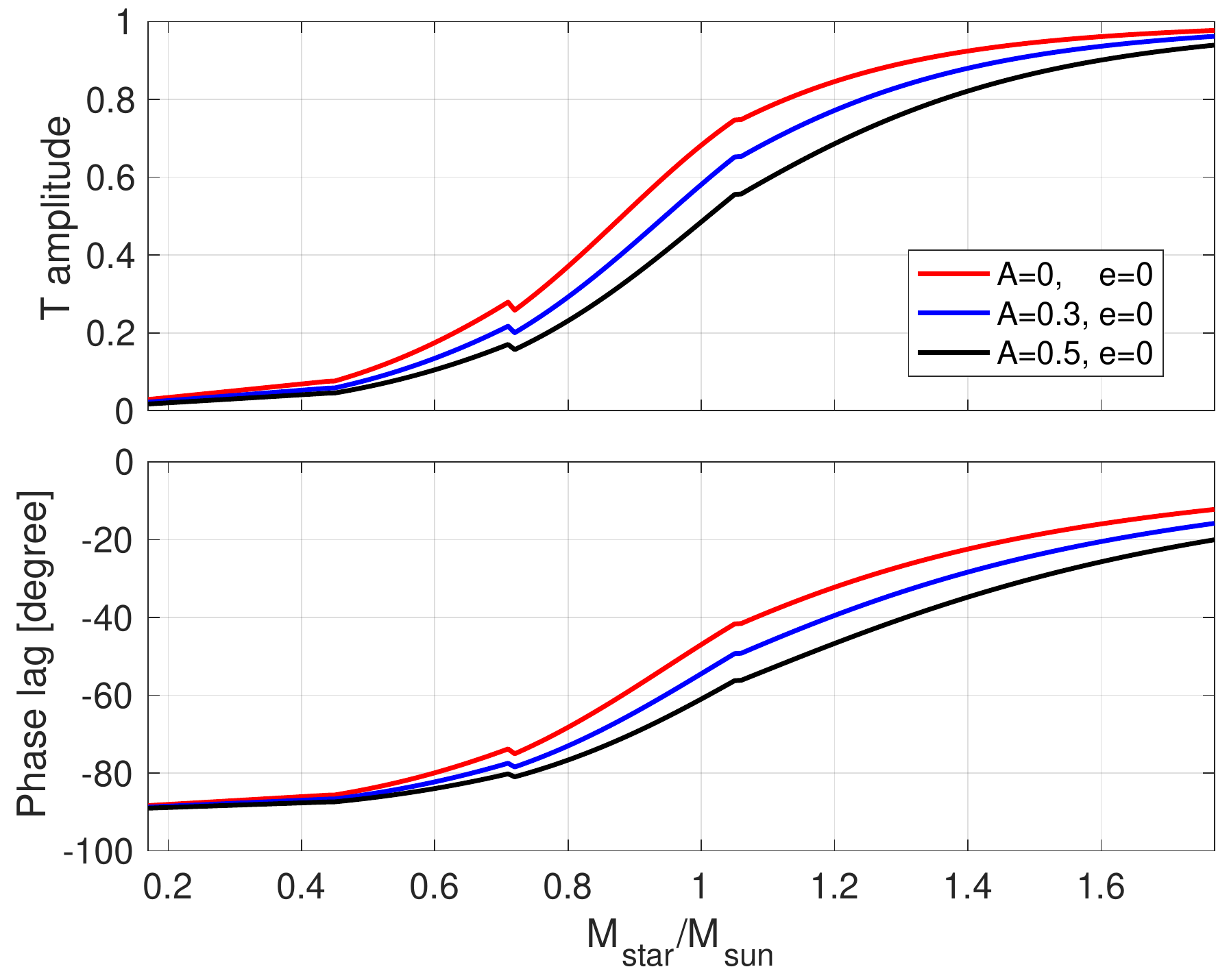}
\caption{ Relative amplitude (upper panel) and phase lag (bottom panel) of the temperature anomalies relative to the variation of the radiative equilibrium temperature as a function of stellar mass.   Different bond albedo $A$ and eccentricity $e$ are assumed for different curves as labelled. The infrared opacity $\kappa$ is assumed to be $5\times10^{-4}\;{\rm m^2kg^{-1}}$ and the heat capacity $c_p$ is assumed to be $13000 \;{\rm Jkg^{-1}K^{-1}}$.
}
\label{fig.compare}
\end{figure} 

A small $P_{\rm orbit}/\trad$   leads to a small amplitude  and a significant phase lag of the temperature variations relative to the irradiation.   At a minimum level of complexity, we  consider the response of the local  temperature to time varying irradiation crudely as a relaxation to the radiative equilibrium over the radiative timescale as 
\begin{equation}
    \frac{dT}{dt}=\frac{T_{\rm RE}(t)-T}{\trad},
\end{equation}
where  $T_{\rm RE}=\overline{T}_{\rm RE}+\Delta T_{\rm RE} \cos{\omega t}$ is the time-varying radiative equilibrium temperature, $\overline{T}_{\rm RE}$ is the time-independent component and $\omega=2\pi/P_{\rm orbit}$. The time-dependent temperature is 
\begin{equation}
    T=\overline{T}_{\rm RE}+\frac{\Delta T_{\rm RE}}{\sqrt{1+\omega^2\trad^2}}\cos{(\omega t+\alpha)},
\end{equation}
where $\sin{\alpha}=-\frac{\omega \trad}{\sqrt{1+\omega^2\trad^2}}$ and $\alpha$ is a phase lag of the temperature perturbation relative to the irradiation. In the presence of a finite $\trad$, temperature varies with an amplitude a factor of $1/\sqrt{1+\omega^2\trad^2}$ of that of the radiative equilibrium, and the phase lag $\alpha$  is in between $-90^{\circ}$ and 0.  A larger $P_{\rm orbit}/\trad$ leads to a stronger temperature variation and  a smaller magnitude of phase lags $\alpha$. Figure \ref{fig.compare} shows the amplitude as a function of stellar mass  in the upper panel and the phase lag in the lower panel assuming $e=0$,  $\kappa_{\rm ref}$, $c_{p,{\rm ref}}$ and a few values for $A$. It appears that the seasonal temperature variations are negligible and there is a nearly $-90^{\circ}$ phase lag near the photosphere for Jupiter-like planets around stars with masses $ \lessapprox 0.6\;M_{\odot}$.

\section{Radiative and temperature responses to time varying irradiation}
\label{ch.rt}
The timescale comparison has laid out the essence of our argument. In the following we take a step further and present one-dimensional (1D) analytic and two-dimensional (2D) numerical  calculations of radiative transfer and atmospheric thermal structure responding to the time-varying irradiation. These calculations demonstrate some quantitative differences  to those calculated by simple relaxation, but the qualitative conclusions remain the same. In addition, the radiative transfer models can capture the pressure-dependent seasonal variations in a more natural way.

{\btt In Section \ref{ch.semigrey} and \ref{ch.nongrey},  we perform 1D analytic calculations in which  we do not attempt to realistically model the geometry and magnitude of the seasonal irradiation  but merely  capture the characteristic  response to a  periodically varying irradiation. Therefore,  the models are linearized in which the response is normalized to the  magnitude of the irradiation variation.} One may consider the calculations as for a latitudinal band  of a rapidly rotating planets (such that the diurnal cycle can be negligible) with non-zero obliquity.

{\btt Seasonality at certain latitudes depends critically on the   magnitude of the seasonal variation of irradiation which depends  on the planetary obliquity, eccentricity and the latitude. In many cases, the seasonal variations are not  sinusoidal. To capture the geometry and magnitude of the seasonal variations, we carry out 2D numerical calculations as a function of latitude and pressure in Section \ref{ch.numerical}. }

\subsection{An analytic semi-grey model}
\label{ch.semigrey}
We start with a semi-grey framework in which  the stellar irradiation  is represented by a single visible band with a characteristic constant opacity and the atmospheric thermal IR fluxes are represented with  another single band with a characteristic constant IR opacity.  Here
we consider a plane-parallel, two-stream approximation   which has been widely used in modeling planetary atmospheres. 
With a steady irradiation and boundary conditions, the structure at  radiative equilibrium  can be straightforwardly found (e.g., \citealp{guillot2010,heng2012effects} in the context of hot Jupiters).  We do not  concern the equilibrium  structure  but are interested in the perturbations  around the  equilibrium. {\btt The variation of irradiation is parameterized as a sinusoidally varying flux perturbation  with an arbitrary amplitude assuming zero eccentricity.} 
The equation set and analytic solution of perturbations of temperature and thermal flux  are given in the  Appendix \ref{ch.appendix1}. 

{\btt The seasonal variations of irradiation are not sinusoidal in many cases. In the presence of a non-zero eccentricity and obliquity, the seasonal variation at any latitude is not sinusoidal; even with a circular orbit, the variation is not sinusoidal at some latitudes. However, the seasonal variation of irradiation may be approximated as a linear combination of sinusoids with different frequencies and amplitudes. The thermal response   can then be the linear combination of solutions to these sinusoids because of the linearity of our problem.    }

One  may express  perturbations of the top-of-atmosphere (TOA) flux $F'(\tau=0,t)$ and photospheric temperature $T'(\tau=1,t)$ in the form of
\begin{equation}
    \begin{split}
        \frac{F'(\tau=0,t)}{\Delta F_v} = \frac{|F'|}{\Delta F_v}\exp(i\alpha_f)\exp(i\omega t), \\ 
        T'(\tau=1,t)=|T'(\tau=1)|\exp(i\alpha_T)\exp(i\omega t),
    \end{split}
    \label{eq.waveform}
\end{equation}
where $\Delta F_v$ is the magnitude of the irradiation variation. Physical quantities are the real parts of Eq. (\ref{eq.waveform}).
The relative amplitude $|F'|/\Delta F_v$ and phase lag $\alpha_f$ of the TOA thermal flux perturbation, as well as the phase lag of the temperature perturbation $\alpha_T$  are determined only by $P_{\rm orbit}/\trad$. The amplitude of temperature perturbations   $|T'|$ depends on $T_{\rm eq}$ and $\Delta F_v$ too, but the ratio to a reference amplitude $|T'|/|T'_{\rm ref}|$ depends on $P_{\rm orbit}/\trad$ only, where $T'_{\rm ref}$ is a perturbation with an extremely large $P_{\rm orbit}/\trad$ which is arbitrarily set to $10^4$. 

\begin{figure}
    \epsscale{1.1}
    \centering
    \plotone{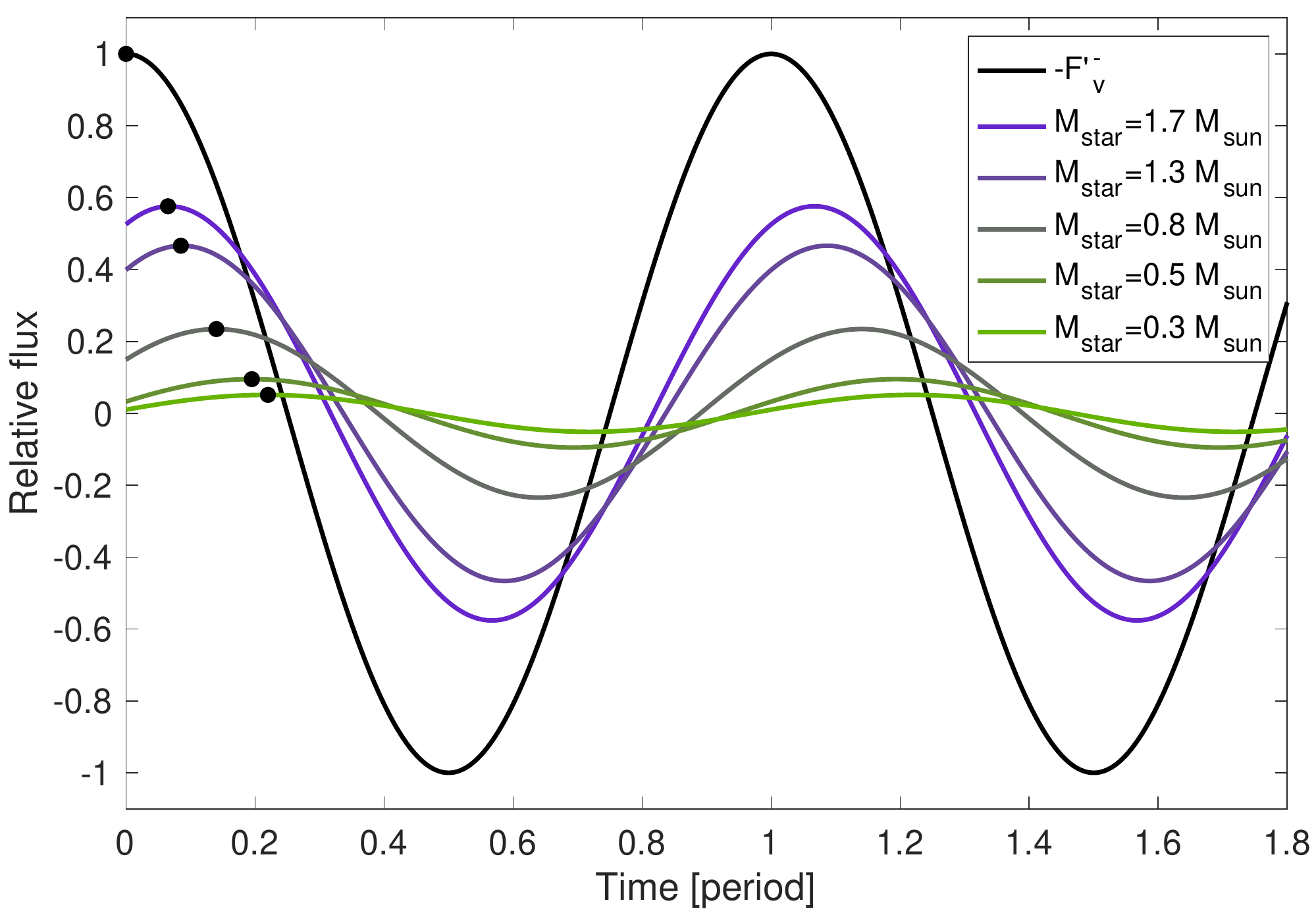}
    \caption{Time evolution of seasonal irradiation (with a reversed sign) and TOA thermal flux variation for cases with different stellar mass. The dots mark the peak fluxes at their first period and illustrate phase lags between different cases. Parameters used in these cases are $\kappa=5\times10^{-4}\;{\rm m^2kg^{-1}}$, $c_p=13000 \;{\rm Jkg^{-1}K^{-1}}$, $\kappa_v=2\times10^{-4}\;{\rm m^2kg^{-1}}$, $\mu_v=1$, $A=0.5$ and $e=0$.}
    \label{fig.fluxtime}
\end{figure}

\begin{figure*}      
\epsscale{1}      
\centering
\plottwo{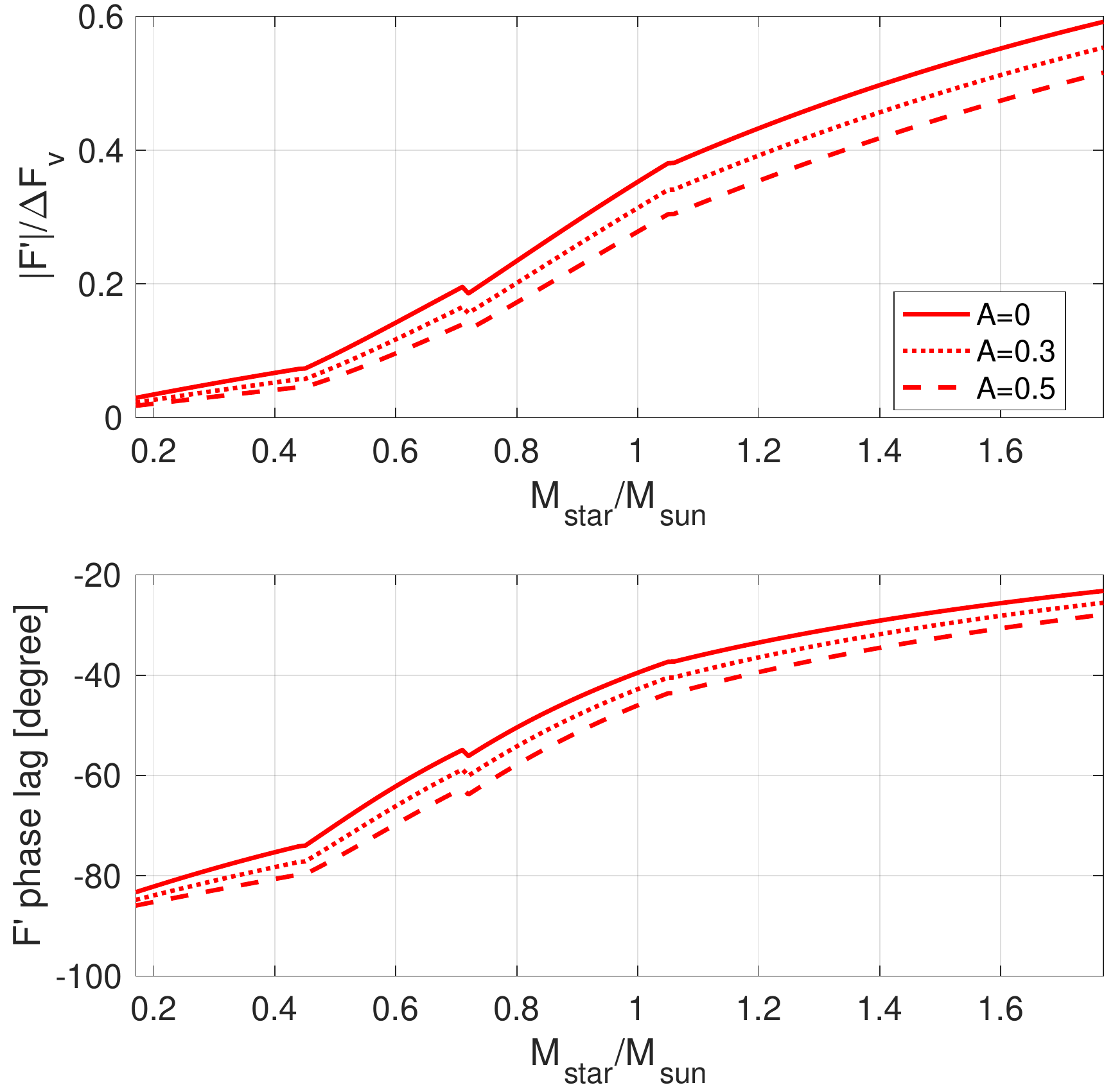}{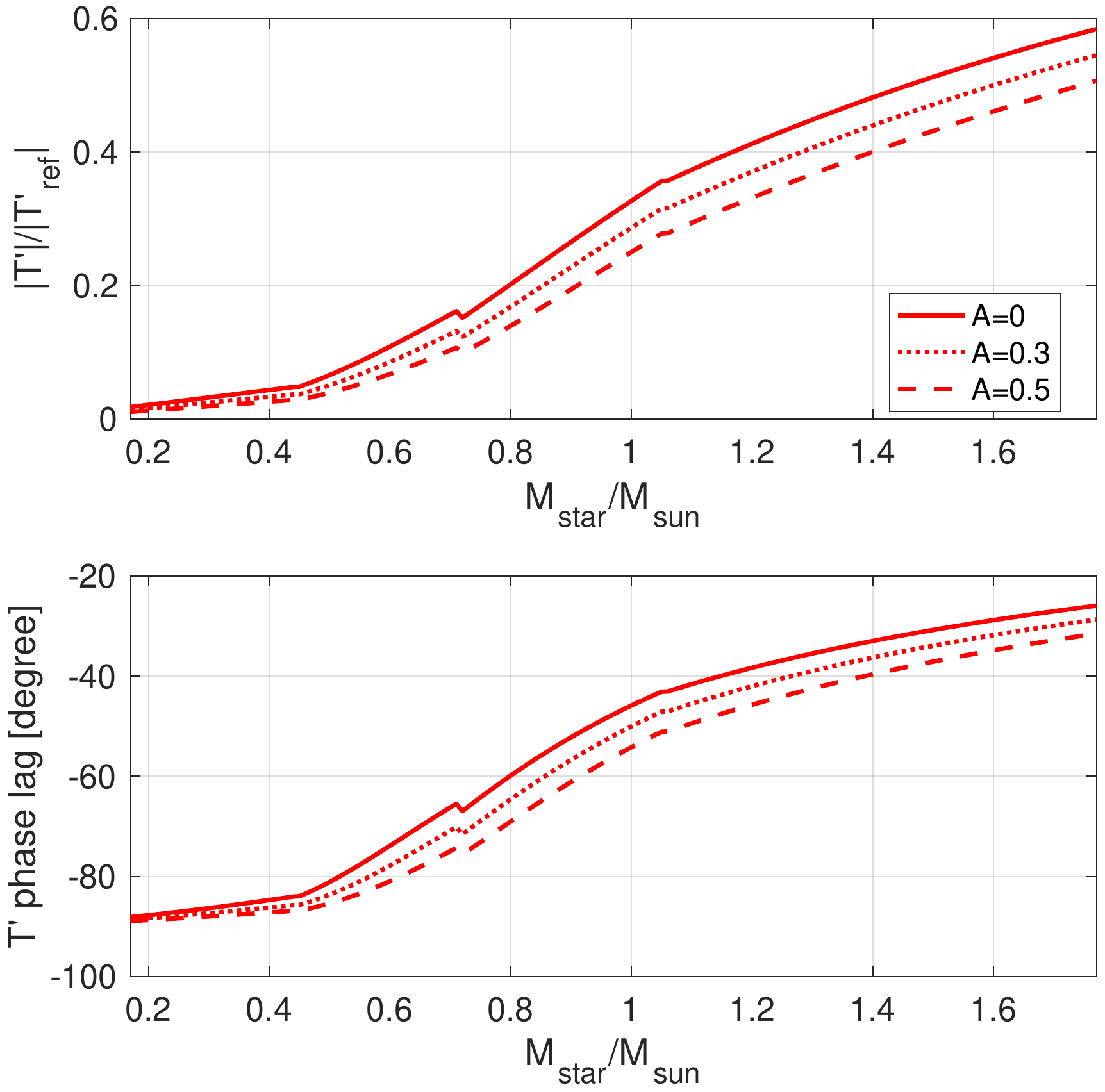}
\caption{\emph{Left:} Relative amplitude (upper panel) and phase lag (bottom panel) of the top-of-atmosphere (TOA) thermal flux anomalies relative to the variation of the irradiation as a function of stellar mass. Different bond albedo $A$  are assumed for different curves, and eccentricity is assumed to be zero for all models. The infrared opacity $\kappa$ is assumed to be $5\times10^{-4}\;{\rm m^2kg^{-1}}$ and the heat capacity $c_p$ is assumed to be $13000 \;{\rm Jkg^{-1}K^{-1}}$. \emph{Right}: Relative amplitude   of the temperature anomalies at optical depth of one relative to that calculated using a large ratio of $P_{\rm orbit}/\trad=10^4$ in the upper panel. The bottom panel shows the phase lag of temperature anomalies at $\tau=1$ relative to the irradiation anomaly.
}
\label{fig.flux}
\end{figure*}

As an example, we assume $\kappa=\kappa_{\rm ref}$, $c_p=c_{p,{\rm ref}}$, visible opacity $\kappa_v=2\times10^{-4}\;{\rm m^2kg^{-1}}$, zenith angle of the irradiation $\mu_v=1$, $A=0.5$ and $e=0$. Time evolution of the seasonal irradiation (with a reversed sign) and the TOA thermal flux variations for cases with different stellar mass are shown in Figure \ref{fig.fluxtime}.  Both the flux magnitude  and the phase lag decrease with decreasing stellar mass. {\btt More results of relative amplitudes and phase lags for thermal flux and temperature perturbations are shown in Figure \ref{fig.flux} as a function of stellar mass for cases with a few different  bond albedos.} The relative amplitudes and phase lags of both $F'(\tau=0)$ and $T'(\tau=1)$ as a function of stellar mass exhibit the same general trend  as that in Section \ref{ch.timescale}. These relative amplitudes  are $\lessapprox 0.1$ at low stellar mass ($\lessapprox 0.6\;M_{\odot}$), quantitatively similar to those using relaxation method (Figure \ref{fig.compare}). However, they increase  only up to  $\lesssim 0.6$ at $M_{\star} = 1.7\;M_{\odot}$, and this upper bound is obviously smaller than that (above 0.95) in the relaxation calculation. All phase lags from  two types of models are nearly $-90^{\circ}$ at $M_{\star}<0.2\;M_{\odot}$. The phase lags in the radiative transfer calculation increase (i.e., the amplitudes  decrease) more rapidly with increasing $M_{\star}$ when $M_{\star}\lessapprox 1\;M_{\odot}$ than that in the relaxation calculation, but their upper bound is lower at the highest $M_{\star}$.    The difference between results from the radiative transfer and relaxation calculations at high stellar mass  presumably arises  from the fact that  communications between the deeper layers where there is a  growing thermal inertia and the photosphere  is naturally taken into account in the radiative transfer calculation while not in the relaxation calculation. 

{\bttt The mean visible opacity is usually smaller than the mean IR opacity in temperate atmospheres \citep{freedman2014}. However,  enhanced absorption of irradiation may be possible. We have calculated cases with $\kappa_v>\kappa$. The relative amplitudes and phase lags of the TOA thermal flux variations are generally larger than those shown in Figure \ref{fig.flux}. This is reasonable because in this case energy is absorbed and re-emitted at a lower thermal optical depth where the thermal inertia is smaller.   }

{\bttt We have focused on the TOA flux and temperature variations near the thermal  photosphere. The amplitude of the perturbations decreases exponentially with increasing optical depth. They are difficult to observe and may be easily affected by other sources such as atmospheric dynamics or interior convection. Thus, the deep perturbations in our calculations are not discussed in the main text but are  referred to Appendix \ref{ch.appendix1} for  details.  }

\subsection{An analytic non-grey model: picket fence in the thermal emission and arbitrary bands in the visible absorption}
\label{ch.nongrey}

We are also interested in the seasonal responses  at different layers  because these might provide observational signatures to better constrain planetary seasonality.   Non-grey (wavelength dependent) effects in the absorption and emission  affect the radiative heating and cooling rates and therefore  the amplitudes and  phase lags at different layers. To capture the non-grey effects in the upper atmospheric layers, we consider a model with the picket fence opacity structure in the thermal band and arbitrary bands in the visible. This setup enables improvements in the radiative transfer calculation while   remaining analytically trackable. We still consider absorption only. 

The picket fence model considers two different values for the thermal opacity  $\kappa_1$ and $\kappa_2$, and  the characteristic  band width associated with them are assumed to be much smaller than the width over which the Planck function varies. The Planck function  is considered constant over a characteristic band width that contains both $\kappa_1$ and $\kappa_2$. Therefore, the total (wavelength-integrated) thermal emission through bands with $\kappa_1$ and $\kappa_2$ are approximated as $(1-\chi)B$ and $\chi B$, respectively, where $\chi$ is the fractional band width associated with $\kappa_2$. For more detailed descriptions of the picket fence models, see \cite{Chandrasekhar1935,king1955} and \cite{parmentier2014a}. In the visible absorption, we consider partitioning of irradiative energy into multiple channels with different opacities. The radiative transfer equations and analytic solution are given in the Appendix \ref{ch.appendix}.

\begin{figure}      
\epsscale{1.13}      
\centering
\plotone{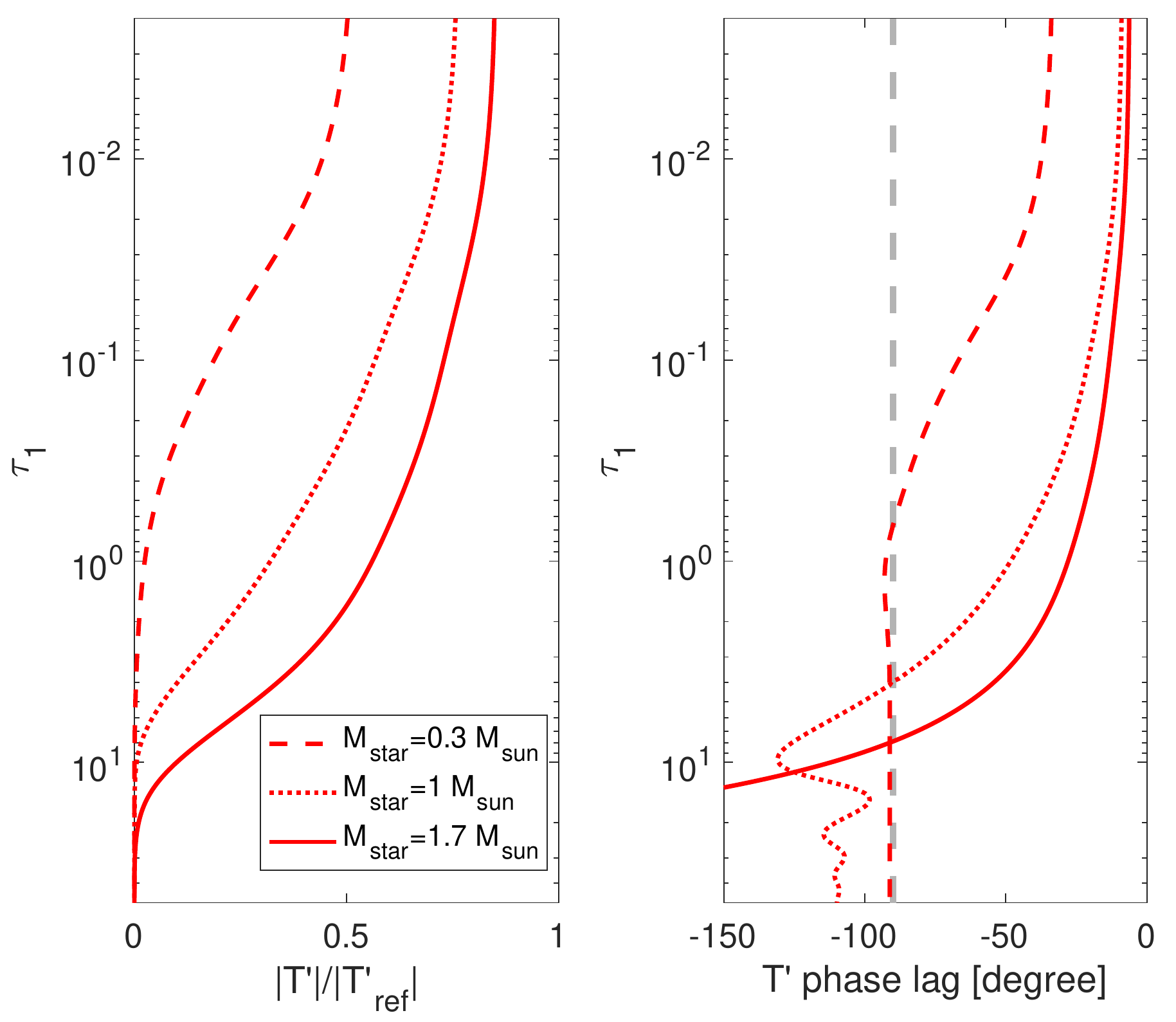}
\caption{\emph{Left:} Relative amplitudes  of the temperature variations relative to that calculated using a large ratio of $P_{\rm orbit}/\trad=10^4$ as a function of optical depth from a   non-grey model with three different stellar mass. These calculations assume infrared opacities of $\kappa_1=5\times10^{-4}\;\mmkg$, $\kappa_2=5\times10^{-3}\;\mmkg$, $\chi=0.3$ and three visible bands with $\kappa_v=[2\times10^{-4},2\times10^{-3},2\times10^{-2}]\;\mmkg$ and $\epsilon=[0.7, 0.2, 0.1]$. The bond albedo $A$ and the eccentricity $e$ are assumed zero. \emph{Right:} The phase lags of the temperature variations relative to the irradiation variation as a function of optical depth for  cases shown on the left panel. {\bttt The grey dashed line represents a $-90^{\circ}$ phase lag.}
}
\label{fig.nongreycase}
\end{figure}

We perform an example  to show the optical-depth dependent response. In principle, the values of $\kappa_1$, $\kappa_2$ and visible opacities may be chosen according to the full opacities for specific cases with certain {\btt atmospheric temperature, composition, and stellar spectrum} \citep{parmentier2015,lee2021}. Here we are interested in a general conceptual illustration, and so in the following results we do not tune the opacities to match specific cases. 
Figure \ref{fig.nongreycase} shows relative amplitudes on the left and phase lags on the right for temperature variations as a function of optical depth for a set of  cases with different stellar mass and  with $\kappa_1=5\times10^{-4}\;\mmkg$, $\kappa_2=5\times10^{-3}\;\mmkg$, $\chi=0.3$ and three visible bands with $\kappa_v=[2\times10^{-4},2\times10^{-3},2\times10^{-2}]\;\mmkg$ and partition coefficients of visible bands $\epsilon=[0.7, 0.2, 0.1]$. As expected, the temperature response is weak at high optical depth and smoothly increases with decreasing $\tau_1$. Planets around higher mass stars have overall larger temperature perturbations due to the longer orbital period compared to the radiative timescale.  {\bttt The phase lags of all cases are relatively large at low optical depth  and decrease with increasing optical depth. These phase lags all go across  $-90^{\circ}$ from low to high $\tau_1$ and this crossing occurs at higher $\tau_1$ for cases with higher stellar mass. After the $-90^{\circ}$ crossing, the phase lag in the case with 0.3 $M_{\odot}$ quickly converges to nearly $-90^{\circ}$ at high optical depth; the one with 1 $M_{\odot}$ first shows  oscillations then converges to a value $<-100^{\circ}$ at depth; the one with 1.7 $M_{\odot}$ continuously  decreases with increasing $\tau_1$. More discussion of the deep perturbations is referred to Appendix \ref{ch.appendix1}.
}

\begin{figure*}      
\epsscale{1}      
\centering
\plottwo{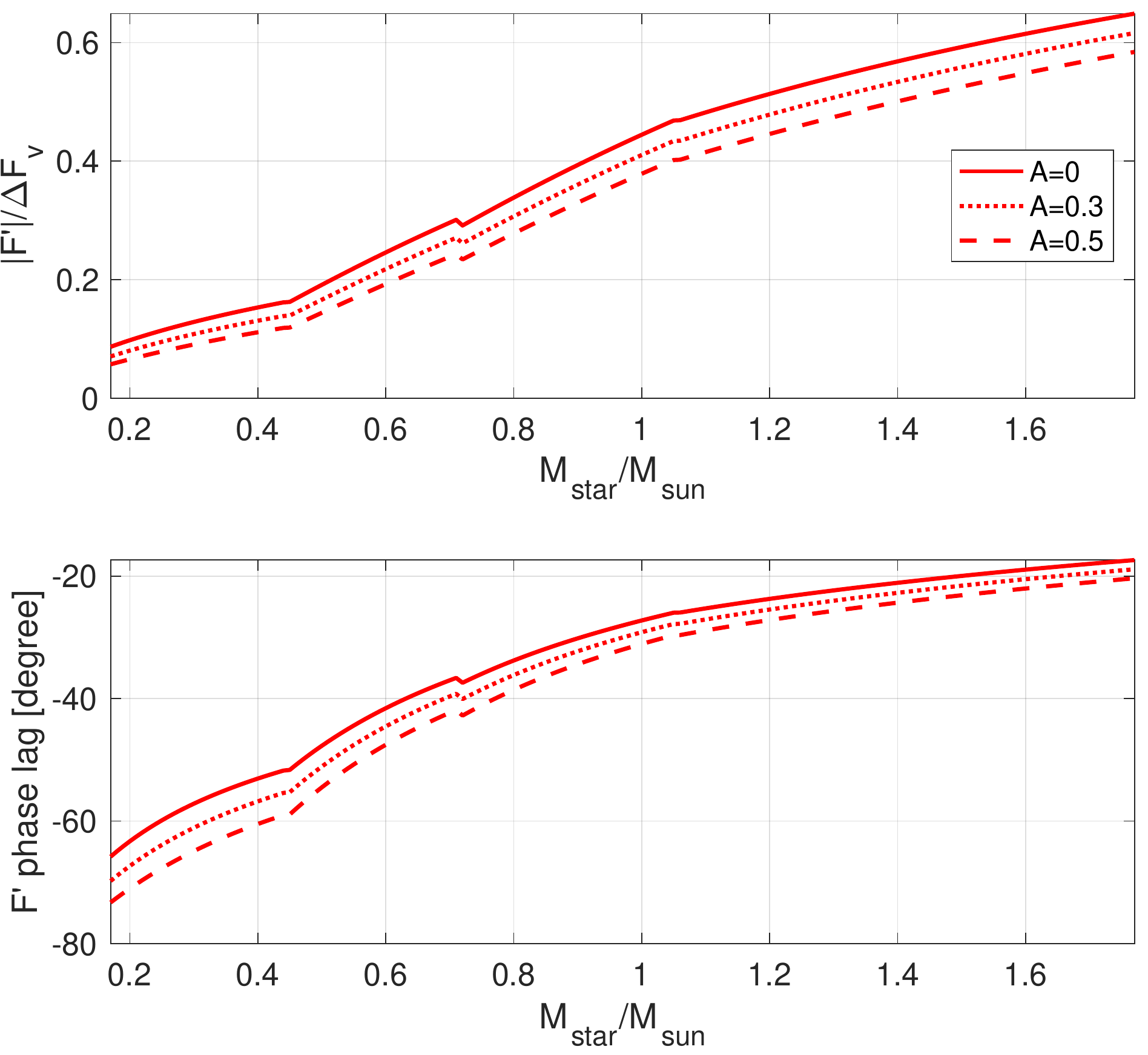}{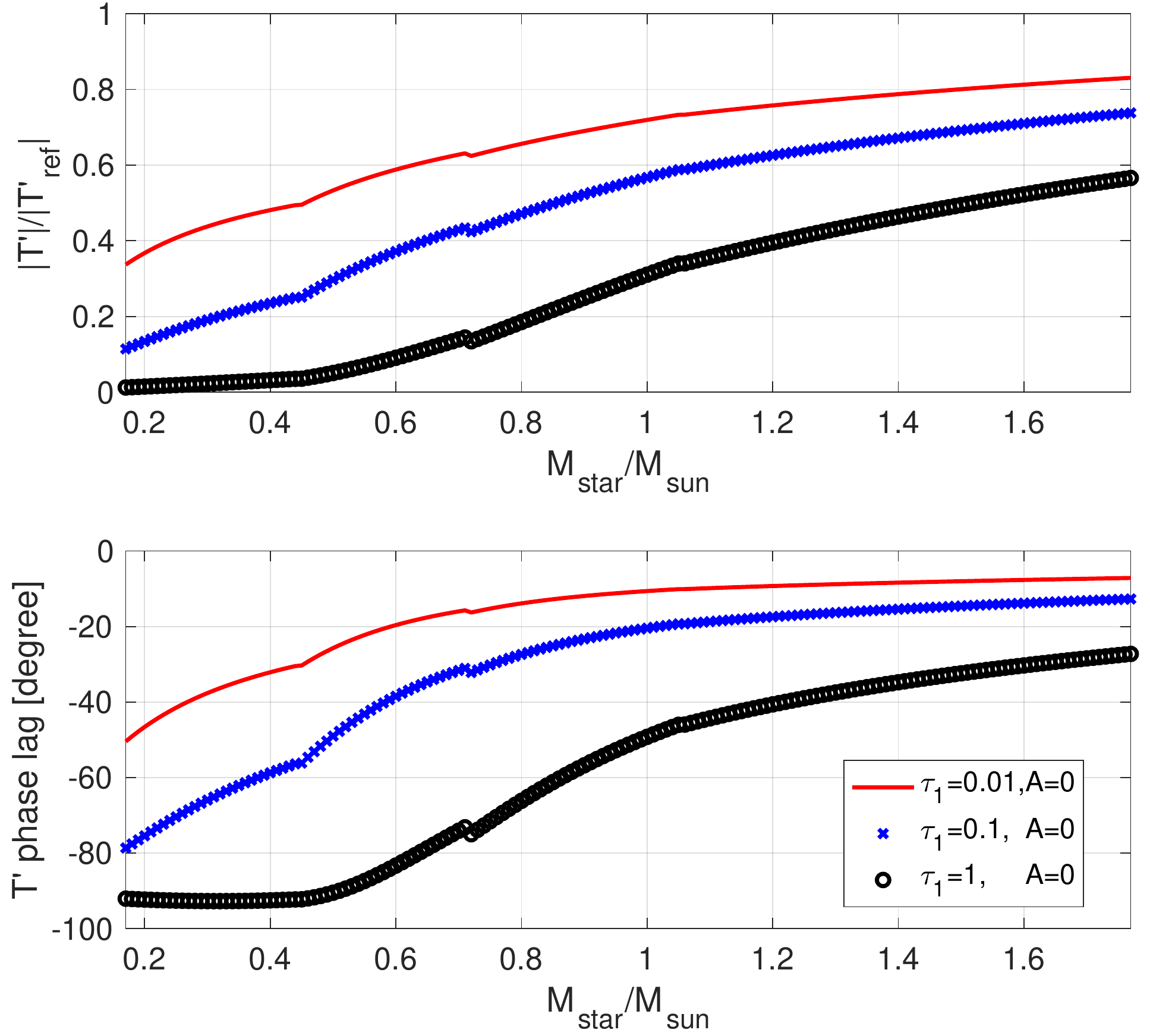}
\caption{\emph{Left:} Relative amplitudes (upper panel) and phase lags (bottom panel) of the TOA thermal flux variations relative to the variation of the irradiation as a function of stellar mass from a set of non-grey cases. Different bond albedo $A$  are assumed for different curves, and the eccentricity is assumed to be zero for all models. Parameters are the same as those used in Figure \ref{fig.nongreycase}. \emph{Right:} Relative amplitudes (upper panel) and phase lags (bottom panel) of temperature variations at three optical depth $\tau_1=0.01, 0.1$ and 1 as a function of stellar mass for the a subset of non-grey cases.
}
\label{fig.nongreyflux}
\end{figure*}

The left panels in Figure \ref{fig.nongreyflux} show  amplitudes and phase lags of TOA thermal fluxes as a function of stellar mass using atmospheric parameters as those used in Figure \ref{fig.nongreycase}. The TOA fluxes show  the same trend as that in the semi-grey case in Figure \ref{fig.flux}, but overall the $|F'|/\Delta F_v$ is slightly higher and the phase lags magnitude is smaller. This is because some portion of the irradiation is absorbed and re-emitted from lower pressures where the responsive timescales are shorter. On the right column, we display temperature perturbation amplitudes and phase lags at different optical depth. At higher stellar masses, the phase lag differences between different optical depth are relatively small ($\lessapprox30^{\circ}$), whereas they show a maximum of about $60^{\circ}$ in cases with the stellar mass of about 0.6 $M_{\odot}$. Then the phase lag differences decline again at lower stellar mass.

Saturn has the most detailed measurements of seasonal variability among gaseous planets  thanks to the longevity of the Cassini mission.  Assuming that the atmospheric parameters are similar to those of Jupiter with $A\sim0.3$,  $P_{\rm orbit}/\trad$ is about 4.6 for Saturn according to Eq. (\ref{eq.essence}), suggesting a somewhat strong seasonality. Indeed, observations and detailed radiative transfer calculations show obvious seasonal hemispheric asymmetry   in Saturn's stratosphere. The amplitudes of temperature variations increase with decreasing pressure. For instance, the  amplitude of the south pole is about 40 K, 10 K and 4 K at pressure between 1 to 5 mbar, 110 mbar and 440 mbar, respectively. The phase lag is about $-30^{\circ}$ and $-15^{\circ}$ at 1 mbar and 0.1 mbar, respectively. See \cite{fletcher2018,fletcher2020} for comprehensive reviews. Here we do not attempt to model the case for Saturn, but these observations are in agreement to our qualitative argument.

\subsection{\btt Numerical  two-dimensional semi-grey calculations}
\label{ch.numerical}

To capture the geometry and magnitude of the seasonal irradiation at different latitudes, we turn to numerical integration of radiative transfer equations performed over a latitudinal grid. At each latitude, the irradiation is treated as  the longitudinal mean which is reasonable for temperate, rapidaly rotating exoplanets \citep{showman2015,rauscher2017}. 

The stellar irradiation pattern in the presence of non-zero obliquity and eccentricity is described as follow. The latitude of the substellar point $\phi_{\rm ss}$ in our model is given by 
\begin{equation}
    \sin\phi_{\rm ss} = \sin\psi\cos{f},
    \label{eq.ss}
\end{equation}
where $\psi$ is the planet's obliquity and $f$ is the true anomaly. The relation between $f$ and time $t$ is linked by the eccentric anomaly $E$ \citep{murray1999}:
\begin{equation}
    \tan{\frac{f}{2}} = \sqrt{\frac{1+e}{1-e}}\tan{\frac{E}{2}}.
\end{equation}
The time evolution of $E$ is given by the Kepler's equation
\begin{equation}
    \frac{dE}{dt}=\frac{2\pi}{P_{\rm orbit}}\frac{1}{1-e\cos{E}}.
\end{equation}
The star-planet distance is $d=a(1-e\cos{f})$. Equation (\ref{eq.ss}) states that the maximum Northern excursion of the substellar latitude ($\phi_{\rm ss}=\psi$) occurs when the planet is at the periapsis ($f=0$). Of course, Equation (\ref{eq.ss}) is only one of the possible configurations, but here we show this particular case for illustrations of seasonal behaviors in the presence of both non-zero obliquity and eccentricity. 
The longitudinal averaged stellar flux at the top of the atmosphere at certain latitude $\phi$ is (e.g., \citealp{liou2002})
\begin{equation}
\begin{split}
    F_v(\phi,t) = & \frac{(1-A)\sigma T^4_{\star}(R_{\star}/d)^2}{\pi} \times \\
    & (\sin{\phi}\sin{\phi_{\rm ss}} H+\cos{\phi}\cos{\phi_{\rm ss }} \sin{H}),
    \label{eq.Fv2}
\end{split}
\end{equation}
where $H=\cos^{-1}(-\tan{\phi}\tan{\phi_{\rm ss}})$ is the length from star-rise to noon at latitude $\phi$ measured in radiance.  When the argument in the inverse cosine function is larger than 1 or smaller than -1, $H$ is set to 0 or $\pi$, respectively.  Note that $d$, $H$ and $\phi_{\rm ss}$ are functions of time $t$.

The 2D numerical model integrates a set of radiative transfer equations (Equations \ref{eq.Fv2}, \ref{eq.rt} and \ref{eq.dTdt}) as a function of latitude and time. We adopt a thermal opacity of $\kappa=5\times10^{-4}\;{\rm m^2kg^{-1}}$,  a visible opacity of $\kappa_v=2\times10^{-4}\;{\rm m^2kg^{-1}}$, and specific heat $c_p=13000 \;{\rm Jkg^{-1}K^{-1}}$, the same  as those in Section \ref{ch.semigrey}. We utilize the numerical package ${\it twostr}$ \citep{kst1995} to integrate the radiative transfer equations, and the mean zenith angle of irradiation is assumed to be $1/\sqrt{3}$ following the default value in ${\it twostr}$.  The model is initialized with a uniform temperature-pressure profile that is in a radiative equilibrium with respect to the global-annual-mean irradiation. At the lower boundary layer, we apply a net upward heat flux with an internal temperature  of 100 K, appropriate for Jupiter-like planets. The simulations are usually integrated for over $10^5$ Earth days after which they are in  statistical equilibrium. 

\begin{figure*}      
\epsscale{1.15}      
\centering
\plotone{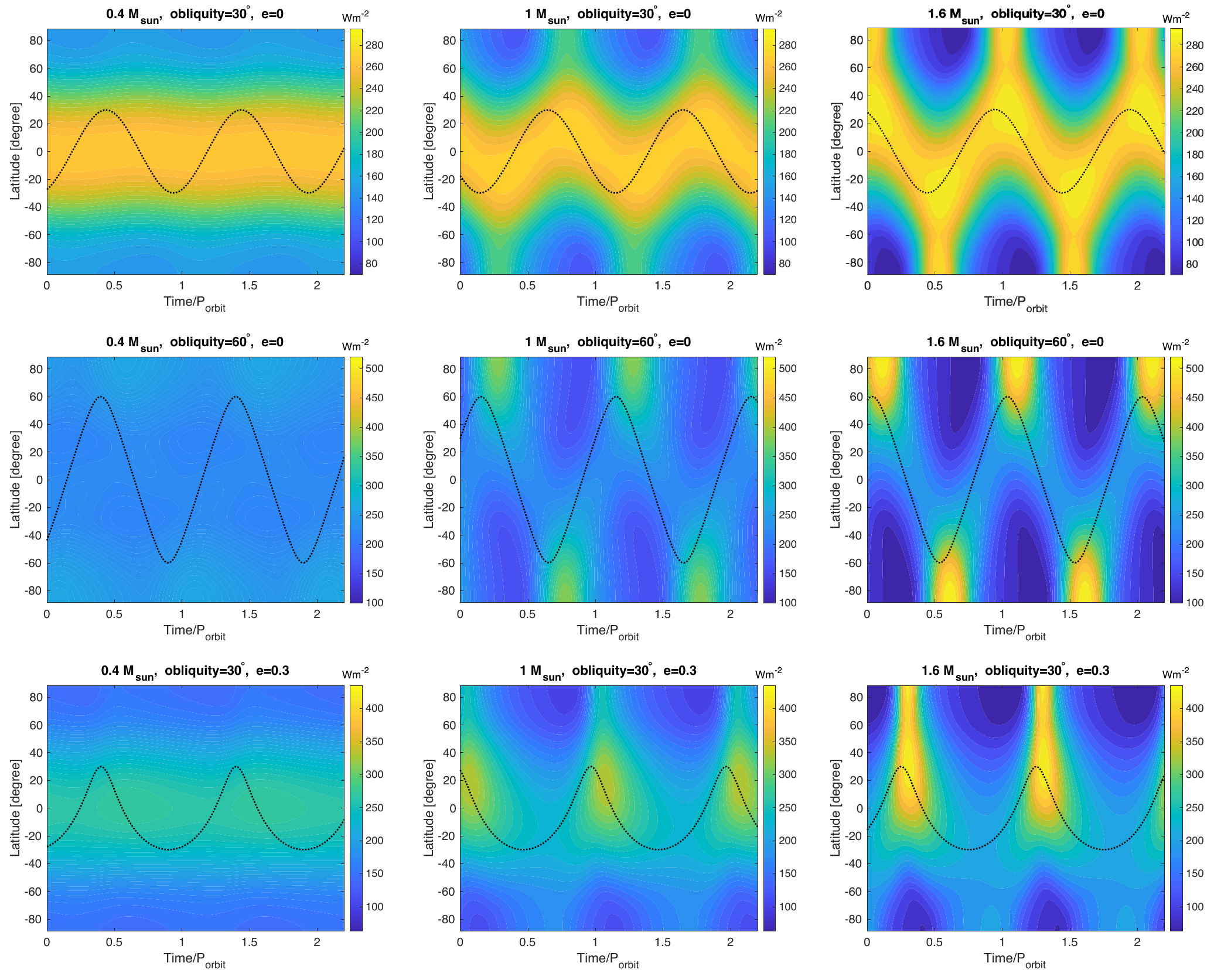}
\caption{Top-of-atmosphere thermal flux as a function of latitude and time. The dash line in each panel is the substellar latitude as a function of time. All models assume an equilibrium temperature of 250 K, a bond albedo $A=0.3$, a thermal opacity of $\kappa=5\times10^{-4}\;{\rm m^2kg^{-1}}$, a visible opacity of  $\kappa_v=2\times10^{-4}\;{\rm m^2kg^{-1}}$ and the specific heat of $c_p=13000 \;{\rm Jkg^{-1}K^{-1}}$. These models have been integrated over $10^5$ Earth days and have been in statistical equilibrium. \emph{Top row:} results of cases with an obliquity of $30^{\circ}$, a zero eccentricity but with three stellar masses of 0.4 (left), 1 (middle) and 1.6 $M_{\odot}$ (right).  The orbital periods are about 28, 381 and 1427 Earth days, respectively. The stellar effective temperatures are 3564 K  and 7001 K, and the stellar radius are   0.34 and  1.9 $R_{\odot}$ for cases with $M_{\star}=0.4$ and 1.6 $M_{\odot}$, respectively. \emph{Middle row}: same as those at the top row except that the obliquity is $60^{\circ}$. \emph{Bottom row:} same as thoes at the top row except that the eccentricity is $e=0.3$. Accordingly, the orbital periods are about 33, 450 and 1687 Earth days for stellar mass of 0.4, 1 and 1.6 $M_{\odot}$, respectively.
}
\label{fig.ecc}
\end{figure*}

We first show results for planets with a circular orbit ($e=0)$, a prescribed bond albedo $A=0.3$ and a fixed equilibrium temperature $T_{\rm eq}=250$ K, but with different obliquity and around stars with different masses. The top and middle rows in Figure \ref{fig.ecc} show the top-of-atmosphere (TOA) thermal flux as a function of time and latitude for cases with stellar masses of $M_{\star}=0.4$, 1 and 1.6 $M_{\odot}$ and two obliquities of $30^{\circ}$ and $60^{\circ}$. The stellar effective temperature $T_{\star}$ and radius $R_{\star}$ are found using relations in \cite{eker2018}. With the given $T_{\rm eq}$, the orbital periods for  cases with 0.4, 1 and 1.6 $M_{\odot}$ and a zero eccentricity are about 28, 381, 1427 Earth days, respectively, based on Equation (\ref{eq.teq}). In each panel, the dashed line  is the substellar latitude  as a function of time. As is easily visualized, for cases with both obliquities, the TOA thermal flux shows stronger hemispheric asymmetry and higher flux variation over the seasonal cycles around stars with higher masses. Magnitude of the phase lag between the thermal flux and irradiation (represented by the dash lines) increases with decrease stellar mass.  The case with 0.4 $M_{\odot}$ exhibits negligible seasonal variation and hemispheric asymmetry, maintaining a nearly constant hot-equator and cold-poles configuration over the seasons for the case with an obliquity of $30^{\circ}$. Cases with an obliquity of $60^{\circ}$  are at the regime in which    the annual-mean irradiation at the poles are greater than that at the equator \citep{ward1974}, and therefore the one with 0.4 $M_{\odot}$ and $60^{\circ}$ obliquity shows a nearly constant cold-equator and hot-poles configuration.

We also present similar results with a moderate eccentricity of $e=0.3$ and an obliquity of $30^{\circ}$ in the bottom row of Figure \ref{fig.ecc}.  The time variation of the substellar latitude is not sinusoidal as a result of the non-zero eccentricity. The non-zero eccentricity also amplifies the seasonal variation in the specific configuration given by Equation (\ref{eq.ss}). In   cases with higher stellar masses, the seasonal flux variations are more pulsive than those shown in cases with $e=0$ (first row). Despite the larger amplitude of the seasonal variation of irradiation, the thermal flux in the case with 0.4 $M_{\odot}$ still exhibits a rather weak seasonal variability. Overall, the picture that planets around  lower-mass stars tend to exhibit  weaker seasonal response than those around higher-mass stars holds well in the presence of non-zero eccentricity.

\section{Discussion}
\label{ch.discussion}

In this work,  we have shown that  given a set of atmospheric parameters of gaseous planets, the atmospheric response to the seasonal variation of stellar irradiation due to non-zero planetary obliquity and/or orbital eccentricity primarily depends on the stellar properties. Despite complications could arise from diverse atmospheric conditions, gaseous planets around low-mass stars should exhibit statistically weaker seasonality than those around more massive stars. {\btt We carried out timescale comparisons in Section \ref{ch.timescale} for the essential argument. In Section \ref{ch.rt}, we showed further demonstrations using both analytic and numerical calculations of radiative transfer. } This finding may help interpret future  atmospheric characterization of exoplanets around different types of stars via directly imaging missions. 

There are interesting implications for the atmospheric general circulation of planets in systems of different stellar mass. {\btt  Using 2D shallow-water dynamical models, \cite{ohno2019} summarized five circulation regimes depending on the planetary obliquity and the atmospheric radiative timescale $\trad$ (see their Figure 1). In this study, we show that their regime IV and V  with $\trad\gg P_{\rm orbit}$ can mostly apply to gaseous planets around low-mass stars, while their regime II and III with $\trad \lesssim P_{\rm orbit}$ likely apply to those around more massive stars. As such,  our work is complementary to that in \cite{ohno2019}.   } 

{\btt A critical obliquity of  $\psi_c\approxeq 54^{\circ}$ is well-known, below which the annual-mean irradiation is maximized at the equator and above which the annual-mean irradiation is maximized at the poles \citep{ward1974}. It also suggests that, the closer the planetary obliquity to  $\psi_c$, the smaller the meridional gradient of the longitudinal-annual-mean irradiation.  Planets  with an obliquity sufficiently close to the critical value $\psi_c$ around low-mass stars may result in not only weak seasonal variation but also small horizontal temperature variation.  In this regime, the driving force of the global circulation that is  solely raised  by the irradiation difference can be weak. Other forms of the driving force, such as that powered   by convective perturbations \citep{williams1978,showman2019} could potentially shape the circulation and drive multiple strong zonal jets. Planets  with an obliquity sufficiently close to $\psi_c$ but around more massive stars could host significant  seasonal hemispheric asymmetry of the temperature structure (regimes II and III in \citealp{ohno2019}). The circulation patterns should be constrained by the temperature difference and exhibit seasonal hemispheric asymmetry. This circulation transition of planets with obliquities close to $\psi_c$ around moderate-mass stars to low-mass stars would be interesting to investigate. On the other hand,   planets with  obliquities much different from $\psi_c$ around stars of any mass could sustain   large meridional temperature  gradients, regardless whether the temperature patterns vary strongly over the season.  Their atmospheric  circulation  should be largely shaped by the differential irradiation and their seasonal variations as shown in \cite{ohno2019}.}

Planetary rotation plays a key role in the circulation as well.  Near low latitudes, horizontally propagating gravity waves and thermally driven overturning circulations (analog to the Hadley circulation on Earth) could be  efficient to remove temperature anomalies. The latitudinal width of this tropical zone is sensitive to rotation, and  is expected to the wider for slower rotators \citep{held1980}. At mid-to-high latitudes where rotation dominates the large-scale dynamics, meridional heat transport efficiency is expected to be weaker than that at low latitudes and decreases with increasing rotation rate \citep{kaspi2015}. We expect that the meridional temperature structure of rapid rotators may be able to hold closer to that determined purely by seasonal irradiation, while atmospheric dynamics of slow rotators might shape the seasonality further in addition to that determined just by the radiative response.
These may have consequence on cloud formation, haze and chemistry transport, which could have effects on reflective lights probed by  direct imaging of exoplanets.

Our analytic models  are vastly idealized, and multiple factors could introduce complications. First, as mentioned above, atmospheric dynamics can transport heat horizontally and vertically. For rapidly rotating planets such as Saturn, dynamical effects on the seasonal pattern is not dominant \citep{fletcher2020}. But this can be important for slow rotators. 
Second, as alluded in Section \ref{ch.timescale}, variations of opacity due to varying atmospheric metallicity and temperature  contribute to the scattering on the simple trend predicted for hydrogen-dominated atmospheres.
Third, the presence and variations of clouds and hazes represent another major uncertainty in the atmospheres. They contribute to the energy budget of both visible and thermal bands, and sometimes can be dominant. Their coupling to the atmospheric dynamics makes quantitative predictions more challenging. 
Finally, the presence of the internal heat in the gaseous planets when the internal heat is comparable or greater than the irradiation could disrupt the trend predicted in this study. The onset of convection could eliminate the seasonal patterns, which is the case for Saturn \citep{fletcher2018,fletcher2020}.  Another example is Jupiter. Despite that there is an equator-to-pole insolation difference, its thermal emission does not show an equator-to-pole pattern which may be attributed to the interior convection of Jupiter \citep{ingersoll1978}.

Future work should include quantitative predictions of seasonal dynamics for planets around different stellar types using more realistic  models. {\btt  Such models should  include global-scale three-dimensional dynamics which can capture vertically propagating waves, baroclinic instability and their  interactions with the jet streams, as well as proper representation of radiative transfer that represents pressure dependent thermal damping. Parameterization of chemical reactions and haze formation, their dynamical transport and effects on the energy budget are interesting to  include. Post-processing using realistic radiative transfer would be necessary to identify observational signatures of the seasonal dynamics and would be useful for future observational interpretations and strategies. One important aspect is that the observational signature of seasonal variation should be sensitive to pressure because the radiative timescale varies with pressure. A combination of low- and high-resolution spectroscopy will be a valuable tool for diagnosing seasonal variability as they can measure layers at both relatively high and low pressures. The  study of seasonal responses over the planet population around different types of stars will be somewhat analog to the day-to-night temperature variations of the hot Jupiter population, which similarly invoke comparison of the radiative timescale to other key timescales in the system (\citealp{perezbecker2013, komacek2016,komacek2017}, see also the latest data compilation in \citealp{wong2021}). We expect trends in the planet population but also significant deviations due to  intrinsic complications in their atmospheres. }\\

\acknowledgments
We thank  discussion with A.P. Showman, R.T. Pierrehumbert, V. Parmentier and X. Zhang. {\btt We thank the reviewer for constructive and encouraging comments.} This study is supported by European Research Council Advanced Grant (grant agreement No. 740963/EXOCONDENSE, PI: R.T. Pierrehumbert).

\if\bibinc n
\bibliography{draft}

\begin{thebibliography}{60}
\expandafter\ifx\csname natexlab\endcsname\relax\def\natexlab#1{#1}\fi

\bibitem[{Andrews {et~al.}(1987)Andrews, Holton, \& Leovy}]{andrews1987}
Andrews, D.~G., Holton, J.~R., \& Leovy, C.~B. 1987, Middle atmosphere
  dynamics, Vol.~40 (Academic press)

\bibitem[{Bolcar {et~al.}(2016)Bolcar, Feinberg, France, Rauscher, Redding, \&
  Schiminovich}]{bolcar2016initial}
Bolcar, M.~R., Feinberg, L., France, K., Rauscher, B.~J., Redding, D., \&
  Schiminovich, D. 2016, in Space telescopes and instrumentation 2016: Optical,
  infrared, and millimeter wave, Vol. 9904, International Society for Optics
  and Photonics, 99040J

\bibitem[{{Chandrasekhar}(1935)}]{Chandrasekhar1935}
{Chandrasekhar}, S. 1935, \mnras, 96, 21

\bibitem[{{Conrath} {et~al.}(1989){Conrath}, {Hanel}, \&
  {Samuelson}}]{conrath1989}
{Conrath}, B.~J., {Hanel}, R.~A., \& {Samuelson}, R.~E. Thermal structure and
  heat balance of the outer planets., ed. S.~K. {Atreya}, J.~B. {Pollack}, \&
  M.~S. {Matthews}, 513--538

\bibitem[{Cowan {et~al.}(2013)Cowan, Fuentes, \& Haggard}]{cowan2013}
Cowan, N.~B., Fuentes, P.~A., \& Haggard, H.~M. 2013, Monthly Notices of the
  Royal Astronomical Society, 434, 2465

\bibitem[{Eker {et~al.}(2018)Eker, Bak{\i}{\c{s}}, Bilir, Soydugan, Steer,
  Soydugan, Bak{\i}{\c{s}}, Ali{\c{c}}avu{\c{s}}, Aslan, \& Alpsoy}]{eker2018}
Eker, Z., Bak{\i}{\c{s}}, V., Bilir, S., Soydugan, F., Steer, I., Soydugan, E.,
  Bak{\i}{\c{s}}, H., Ali{\c{c}}avu{\c{s}}, F., Aslan, G., \& Alpsoy, M. 2018,
  Monthly Notices of the Royal Astronomical Society, 479, 5491

\bibitem[{Ferreira {et~al.}(2014)Ferreira, Marshall, O’Gorman, \&
  Seager}]{ferreira2014}
Ferreira, D., Marshall, J., O’Gorman, P.~A., \& Seager, S. 2014, Icarus, 243,
  236

\bibitem[{{Fletcher}(2021)}]{fletcher2021}
{Fletcher}, L.~N. 2021, arXiv e-prints, arXiv:2105.06377

\bibitem[{{Fletcher} {et~al.}(2018){Fletcher}, {Greathouse}, {Guerlet},
  {Moses}, \& {West}}]{fletcher2018}
{Fletcher}, L.~N., {Greathouse}, T.~K., {Guerlet}, S., {Moses}, J.~I., \&
  {West}, R.~A. {Saturn's Seasonally Changing Atmosphere. Thermal Structure
  Composition And Aerosols}, ed. K.~H. {Baines}, F.~M. {Flasar}, N.~{Krupp}, \&
  T.~{Stallard}, 251--294

\bibitem[{{Fletcher} {et~al.}(2020){Fletcher}, {Sromovsky}, {Hue}, {Moses},
  {Guerlet}, {West}, \& {Koskinen}}]{fletcher2020}
{Fletcher}, L.~N., {Sromovsky}, L., {Hue}, V., {Moses}, J.~I., {Guerlet}, S.,
  {West}, R.~A., \& {Koskinen}, T. 2020, arXiv e-prints, arXiv:2012.09288

\bibitem[{Freedman {et~al.}(2014)Freedman, Lustig-Yaeger, Fortney, Lupu,
  Marley, \& Lodders}]{freedman2014}
Freedman, R.~S., Lustig-Yaeger, J., Fortney, J.~J., Lupu, R.~E., Marley, M.~S.,
  \& Lodders, K. 2014, The Astrophysical Journal Supplement Series, 214, 25

\bibitem[{Gaidos \& Williams(2004)}]{gaidos2004}
Gaidos, E. \& Williams, D. 2004, New Astronomy, 10, 67

\bibitem[{Guendelman \& Kaspi(2019)}]{guendelman2019}
Guendelman, I. \& Kaspi, Y. 2019, The Astrophysical Journal, 881, 67

\bibitem[{{Guillot}(2010)}]{guillot2010}
{Guillot}, T. 2010, \aap, 520, A27

\bibitem[{Held \& Hou(1980)}]{held1980}
Held, I.~M. \& Hou, A.~Y. 1980, Journal of the Atmospheric Sciences, 37, 515

\bibitem[{Heng {et~al.}(2012)Heng, Hayek, Pont, \& Sing}]{heng2012effects}
Heng, K., Hayek, W., Pont, F., \& Sing, D.~K. 2012, Monthly Notices of the
  Royal Astronomical Society, 420, 20

\bibitem[{{Ingersoll} \& {Porco}(1978)}]{ingersoll1978}
{Ingersoll}, A.~P. \& {Porco}, C.~C. 1978, \icarus, 35, 27

\bibitem[{{Kang}(2019)}]{kang2019b}
{Kang}, W. 2019, \apjl, 877, L6

\bibitem[{Kang {et~al.}(2019)Kang, Cai, \& Tziperman}]{kang2019}
Kang, W., Cai, M., \& Tziperman, E. 2019, Icarus, 330, 142

\bibitem[{Kaspi \& Showman(2015)}]{kaspi2015}
Kaspi, Y. \& Showman, A.~P. 2015, The Astrophysical Journal, 804, 60

\bibitem[{Kataria {et~al.}(2013)Kataria, Showman, Lewis, Fortney, Marley, \&
  Freedman}]{kataria2013}
Kataria, T., Showman, A.~P., Lewis, N.~K., Fortney, J.~J., Marley, M.~S., \&
  Freedman, R.~S. 2013, The Astrophysical Journal, 767, 76

\bibitem[{Kawahara \& Fujii(2010)}]{kawahara2010}
Kawahara, H. \& Fujii, Y. 2010, The Astrophysical Journal, 720, 1333

\bibitem[{{King}(1955)}]{king1955}
{King}, I. J. I.~F. 1955, \apj, 121, 711

\bibitem[{Komacek \& Showman(2016)}]{komacek2016}
Komacek, T.~D. \& Showman, A.~P. 2016, The Astrophysical Journal, 821, 16

\bibitem[{{Komacek} {et~al.}(2017){Komacek}, {Showman}, \& {Tan}}]{komacek2017}
{Komacek}, T.~D., {Showman}, A.~P., \& {Tan}, X. 2017, \apj, 835, 198

\bibitem[{{Kreidberg} {et~al.}(2014){Kreidberg}, {Bean}, {D{\'e}sert}, {Line},
  {Fortney}, {Madhusudhan}, {Stevenson}, {Showman}, {Charbonneau},
  {McCullough}, {Seager}, {Burrows}, {Henry}, {Williamson}, {Kataria}, \&
  {Homeier}}]{kreidberg2014}
{Kreidberg}, L., {Bean}, J.~L., {D{\'e}sert}, J.-M., {Line}, M.~R., {Fortney},
  J.~J., {Madhusudhan}, N., {Stevenson}, K.~B., {Showman}, A.~P.,
  {Charbonneau}, D., {McCullough}, P.~R., {Seager}, S., {Burrows}, A., {Henry},
  G.~W., {Williamson}, M., {Kataria}, T., \& {Homeier}, D. 2014, \apjl, 793,
  L27

\bibitem[{Kylling {et~al.}(1995)Kylling, Stamnes, \& Tsay}]{kst1995}
Kylling, A., Stamnes, K., \& Tsay, S.-C. 1995, Journal of Atmospheric
  Chemistry, 21, 115

\bibitem[{Langton \& Laughlin(2007)}]{langton2007}
Langton, J. \& Laughlin, G. 2007, The Astrophysical Journal Letters, 657, L113

\bibitem[{Langton \& Laughlin(2008)}]{langton2008}
---. 2008, The Astrophysical Journal, 674, 1106

\bibitem[{{Lee} {et~al.}(2021){Lee}, {Parmentier}, {Hammond}, {Grimm},
  {Kitzmann}, {Tan}, {Tsai}, \& {Pierrehumbert}}]{lee2021}
{Lee}, E. K.~H., {Parmentier}, V., {Hammond}, M., {Grimm}, S.~L., {Kitzmann},
  D., {Tan}, X., {Tsai}, S.-M., \& {Pierrehumbert}, R.~T. 2021, \mnras, 506,
  2695

\bibitem[{Lewis {et~al.}(2014)Lewis, Showman, Fortney, Knutson, \&
  Marley}]{lewis2014}
Lewis, N.~K., Showman, A.~P., Fortney, J.~J., Knutson, H.~A., \& Marley, M.~S.
  2014, The Astrophysical Journal, 795, 150

\bibitem[{Lewis {et~al.}(2010)Lewis, Showman, Fortney, Marley, Freedman, \&
  Lodders}]{lewis2010}
Lewis, N.~K., Showman, A.~P., Fortney, J.~J., Marley, M.~S., Freedman, R.~S.,
  \& Lodders, K. 2010, The Astrophysical Journal, 720, 344

\bibitem[{Line {et~al.}(2021)Line, Brogi, Bean, Gandhi, Zalesky, Parmentier,
  Smith, Mace, Mansfield, Kempton, {et~al.}}]{line2021}
Line, M.~R., Brogi, M., Bean, J.~L., Gandhi, S., Zalesky, J., Parmentier, V.,
  Smith, P., Mace, G.~N., Mansfield, M., Kempton, E. M.-R., {et~al.} 2021,
  Nature, 598, 580

\bibitem[{Liou(2002)}]{liou2002}
Liou, K.-N. 2002, An introduction to atmospheric radiation, Vol.~84 (Academic
  press)

\bibitem[{{Luger} {et~al.}(2021){Luger}, {Foreman-Mackey}, {Hedges}, \&
  {Hogg}}]{luger2021}
{Luger}, R., {Foreman-Mackey}, D., {Hedges}, C., \& {Hogg}, D.~W. 2021, \aj,
  162, 123

\bibitem[{Mayorga {et~al.}(2021)Mayorga, Robinson, Marley, May, \&
  Stevenson}]{mayorga2021}
Mayorga, L., Robinson, T.~D., Marley, M.~S., May, E., \& Stevenson, K.~B. 2021,
  arXiv preprint arXiv:2105.08009

\bibitem[{Mennesson {et~al.}(2016)Mennesson, Gaudi, Seager, Cahoy,
  Domagal-Goldman, Feinberg, Guyon, Kasdin, Marois, Mawet,
  {et~al.}}]{mennesson2016habitable}
Mennesson, B., Gaudi, S., Seager, S., Cahoy, K., Domagal-Goldman, S., Feinberg,
  L., Guyon, O., Kasdin, J., Marois, C., Mawet, D., {et~al.} 2016, in Space
  telescopes and instrumentation 2016: Optical, infrared, and millimeter wave,
  Vol. 9904, International Society for Optics and Photonics, 99040L

\bibitem[{Murray \& Dermott(1999)}]{murray1999}
Murray, C.~D. \& Dermott, S.~F. 1999, Solar system dynamics (Cambridge
  university press)

\bibitem[{{Nixon} {et~al.}(2010){Nixon}, {Achterberg}, {Romani}, {Allen},
  {Zhang}, {Teanby}, {Irwin}, \& {Flasar}}]{nixon2010}
{Nixon}, C.~A., {Achterberg}, R.~K., {Romani}, P.~N., {Allen}, M., {Zhang}, X.,
  {Teanby}, N.~A., {Irwin}, P.~G.~J., \& {Flasar}, F.~M. 2010, \planss, 58,
  1667

\bibitem[{Ohno \& Zhang(2019{\natexlab{a}})}]{ohno2019}
Ohno, K. \& Zhang, X. 2019{\natexlab{a}}, The Astrophysical Journal, 874, 1

\bibitem[{Ohno \& Zhang(2019{\natexlab{b}})}]{ohno2019b}
---. 2019{\natexlab{b}}, The Astrophysical Journal, 874, 2

\bibitem[{Palubski {et~al.}(2020)Palubski, Shields, \& Deitrick}]{palubski2020}
Palubski, I.~Z., Shields, A.~L., \& Deitrick, R. 2020, The Astrophysical
  Journal, 890, 30

\bibitem[{Parmentier \& Guillot(2014)}]{parmentier2014a}
Parmentier, V. \& Guillot, T. 2014, Astronomy \& Astrophysics, 562, A133

\bibitem[{Parmentier {et~al.}(2015)Parmentier, Guillot, Fortney, \&
  Marley}]{parmentier2015}
Parmentier, V., Guillot, T., Fortney, J.~J., \& Marley, M.~S. 2015, Astronomy
  \& Astrophysics, 574, A35

\bibitem[{{Perez-Becker} \& {Showman}(2013)}]{perezbecker2013}
{Perez-Becker}, D. \& {Showman}, A.~P. 2013, \apj, 776, 134

\bibitem[{Rauscher(2017)}]{rauscher2017}
Rauscher, E. 2017, The Astrophysical Journal, 846, 69

\bibitem[{Schwartz \& Cowan(2015)}]{schwartz2015}
Schwartz, J.~C. \& Cowan, N.~B. 2015, Monthly Notices of the Royal Astronomical
  Society, 449, 4192

\bibitem[{Schwartz {et~al.}(2016)Schwartz, Sekowski, Haggard, Pall{\'e}, \&
  Cowan}]{schwartz2016}
Schwartz, J.~C., Sekowski, C., Haggard, H.~M., Pall{\'e}, E., \& Cowan, N.~B.
  2016, Monthly Notices of the Royal Astronomical Society, 457, 926

\bibitem[{Shields {et~al.}(2016)Shields, Barnes, Agol, Charnay, Bitz, \&
  Meadows}]{shields2016}
Shields, A.~L., Barnes, R., Agol, E., Charnay, B., Bitz, C., \& Meadows, V.~S.
  2016, Astrobiology, 16, 443

\bibitem[{{Showman} \& {Guillot}(2002)}]{showman2002}
{Showman}, A.~P. \& {Guillot}, T. 2002, \aap, 385, 166

\bibitem[{Showman {et~al.}(2015)Showman, Lewis, \& Fortney}]{showman2015}
Showman, A.~P., Lewis, N.~K., \& Fortney, J.~J. 2015, The Astrophysical
  Journal, 801, 95

\bibitem[{{Showman} {et~al.}(2019){Showman}, {Tan}, \& {Zhang}}]{showman2019}
{Showman}, A.~P., {Tan}, X., \& {Zhang}, X. 2019, \apj, 883, 4

\bibitem[{Spergel {et~al.}(2013)Spergel, Gehrels, Breckinridge, Donahue,
  Dressler, Gaudi, Greene, Guyon, Hirata, Kalirai, {et~al.}}]{spergel2013wide}
Spergel, D., Gehrels, N., Breckinridge, J., Donahue, M., Dressler, A., Gaudi,
  B., Greene, T., Guyon, O., Hirata, C., Kalirai, J., {et~al.} 2013, arXiv
  preprint arXiv:1305.5422

\bibitem[{{Sromovsky} {et~al.}(1998){Sromovsky}, {Collard}, {Fry}, {Orton},
  {Lemmon}, {Tomasko}, \& {Freedman}}]{Sromovsky1998}
{Sromovsky}, L.~A., {Collard}, A.~D., {Fry}, P.~M., {Orton}, G.~S., {Lemmon},
  M.~T., {Tomasko}, M.~G., \& {Freedman}, R.~S. 1998, \jgr, 103, 22929

\bibitem[{Su \& Lai(2021)}]{su2021}
Su, Y. \& Lai, D. 2021, Monthly Notices of the Royal Astronomical Society, 509,
  3301

\bibitem[{{Ward}(1974)}]{ward1974}
{Ward}, W.~R. 1974, \jgr, 79, 3375

\bibitem[{Williams \& Kasting(1997)}]{williams1997}
Williams, D.~M. \& Kasting, J.~F. 1997, Icarus, 129, 254

\bibitem[{Williams \& Pollard(2002)}]{williams2002}
Williams, D.~M. \& Pollard, D. 2002, International Journal of Astrobiology, 1,
  61

\bibitem[{Williams(1978)}]{williams1978}
Williams, G.~P. 1978, Journal of the Atmospheric Sciences, 35, 1399

\bibitem[{{Wong} {et~al.}(2021){Wong}, {Shporer}, {Zhou}, {Kitzmann},
  {Komacek}, {Tan}, {Tronsgaard}, {Buchhave}, {Vissapragada}, {Greklek-McKeon},
  {Rodriguez}, {Ahlers}, {Quinn}, {Furlan}, {Howell}, {Bieryla}, {Heng},
  {Knutson}, {Collins}, {McLeod}, {Berlind}, {Brown}, {Calkins}, {de Leon},
  {Esparza-Borges}, {Esquerdo}, {Fukui}, {Gan}, {Girardin}, {Gnilka}, {Ikoma},
  {Jensen}, {Kielkopf}, {Kodama}, {Kurita}, {Lester}, {Lewin}, {Marino},
  {Murgas}, {Narita}, {Pall{\'e}}, {Schwarz}, {Stassun}, {Tamura}, {Watanabe},
  {Benneke}, {Ricker}, {Latham}, {Vanderspek}, {Seager}, {Winn}, {Jenkins},
  {Caldwell}, {Fong}, {Huang}, {Mireles}, {Schlieder}, {Shiao}, \& {Noel
  Villase{\~n}or}}]{wong2021}
{Wong}, I., {Shporer}, A., {Zhou}, G., {Kitzmann}, D., {Komacek}, T.~D., {Tan},
  X., {Tronsgaard}, R., {Buchhave}, L.~A., {Vissapragada}, S.,
  {Greklek-McKeon}, M., {Rodriguez}, J.~E., {Ahlers}, J.~P., {Quinn}, S.~N.,
  {Furlan}, E., {Howell}, S.~B., {Bieryla}, A., {Heng}, K., {Knutson}, H.~A.,
  {Collins}, K.~A., {McLeod}, K.~K., {Berlind}, P., {Brown}, P., {Calkins},
  M.~L., {de Leon}, J.~P., {Esparza-Borges}, E., {Esquerdo}, G.~A., {Fukui},
  A., {Gan}, T., {Girardin}, E., {Gnilka}, C.~L., {Ikoma}, M., {Jensen}, E.
  L.~N., {Kielkopf}, J., {Kodama}, T., {Kurita}, S., {Lester}, K.~V., {Lewin},
  P., {Marino}, G., {Murgas}, F., {Narita}, N., {Pall{\'e}}, E., {Schwarz},
  R.~P., {Stassun}, K.~G., {Tamura}, M., {Watanabe}, N., {Benneke}, B.,
  {Ricker}, G.~R., {Latham}, D.~W., {Vanderspek}, R., {Seager}, S., {Winn},
  J.~N., {Jenkins}, J.~M., {Caldwell}, D.~A., {Fong}, W., {Huang}, C.~X.,
  {Mireles}, I., {Schlieder}, J.~E., {Shiao}, B., \& {Noel Villase{\~n}or}, J.
  2021, \aj, 162, 256

\end{thebibliography}


\begin{thebibliography}
\end{thebibliography}
\fi

\if\bibinc y

\fi

\appendix
\section{Analytic solution of semi-grey perturbations}
\label{ch.appendix1}
The  plane-parallel, two-stream approximation of radiative transfer equations for the diffuse, azimuthally averaged IR intensity $I$ are (e.g., \citealp{liou2002} Chapter 4.6, or \citealp{komacek2017} for the exact setup as here):
\begin{equation}
\begin{split}
    \mu\frac{dI^{+}}{d\tau} = I^{+} - B(\tau), \\
    -\mu\frac{dI^{-}}{d\tau} = I^{-} - B(\tau),
    \label{eq.rt}
    \end{split}
\end{equation}
where $I^{+}$ is the upward intensity, $I^{-}$ is the downward intensity in the IR, $\tau$ is IR optical depth, $\mu$ is the mean IR zenith angle associated with a semi-isotropic hemisphere of radiation, and $B(\tau)$ is the Planck function at the local temperature $T(\tau)$. The diffusive thermal fluxes are obtained as $F^{\pm}=\pi I^{\pm}$. We use  the hemi-isotropic closure for the thermal band with $\mu=0.5$, which assumes that the thermal radiation is isotropic in each hemisphere which is reasonable for atmospheres  without much scattering.  The stellar irradiation in the visible band is included simply as  
\begin{equation}
    F^-_v=F_0 (1-A) \exp{\left(-\frac{\tau_v}{\mu_v}\right)},\quad F^+_v=0,
    \label{eq.Fv}
\end{equation}
where $F_0$ is the top-of-atmosphere (TOA) irradiative flux density at a given location with  $\mu_v$ being the mean local zenith angle of the irradiation, and $\tau_v$ is the optical depth in the visible band. The total atmospheric net flux is $F_{\rm tot}=F^+_v-F^-_v+F^+-F^-$. For simplicity, we consider absorption only and explicitly omit scattering. Effects of scattering  in the energy balance is implicitly included as non-zero bond albedo $A$. The response of  temperature  to the time-dependent radiative fluxes is 
\begin{equation}
    \frac{dT}{dt} = \frac{g}{c_p}\frac{dF_{\rm tot}}{dp} = \frac{\kappa}{c_p}\frac{dF_{\rm tot}}{d\tau}.
    \label{eq.dTdt}
\end{equation}

Perturbations of temperature and thermal fluxes relative to the radiative equilibrium as a function of $\tau$ are denoted as $T'$, $F'^+$ and $F'^-$. The perturbation of the Planck function is approximated as $B'=4\sigma T^3_{\rm RE}T'$. We write the perturbation of the irradiation  as the following periodic oscillation:
\begin{equation}
    F'^-_v(t)= \Delta F_v \exp{(i\omega t)} \exp{\left(-\frac{\tau_v}{\mu_v}\right)},
\end{equation}
in which  the change of $\mu_v$ is neglected.  Substituting these perturbations to Eqs. (\ref{eq.rt}), (\ref{eq.Fv}) and (\ref{eq.dTdt}) and subtracting the radiative equilibrium state, we obtain the equation for the perturbation of the net thermal flux $F'=F'^+-F'^-$:
\begin{equation}
\begin{split}
    & \mu^2\frac{\partial^3F'}{\partial \tau^2\partial t}= \frac{\partial F'}{\partial t}- \\
    & 8\pi\sigma T^3_{\rm RE}\mu \frac{\kappa}{c_p}\left(\frac{\partial^2F'}{\partial\tau^2}-\Delta F_v\exp{(i\omega t)}\exp{(-\beta\tau)}\beta^2\right),
    \end{split}
    \label{eq.F}
\end{equation}
where $\beta=\kappa_v/(\kappa\mu_v)$ and $\kappa_v$ is the opacity at the visible band. For analytic simplicity, we have neglected the variation of $T_{\rm RE}$ with $\tau$ in Eq. (\ref{eq.F}) and   adopt $T_{\rm RE}\sim \teq$ below.
Applying  boundary conditions of vanishing perturbations at $\tau\rightarrow\infty$ and $F'^-=0$ at $\tau=0$,  the solution to Eq. (\ref{eq.F}) is 
\begin{equation}
    F'=\left[C\exp{(-k\tau)}+Z\exp{(-\beta\tau)}\right] \Delta F_v\exp{(i\omega t)},
    \label{eq.solution}
\end{equation}
with constants
\begin{equation*}
    \begin{split}
        & C=-Z\frac{k+\beta\mu k}{\beta+\beta\mu k},\quad
        Z=\frac{\beta^2}{\beta^2+i(\mu\beta^2-1/\mu)/\eta}, \\
        & k=(\mu^2+\eta^2)^{-1/4}\mu^{-1/2}\exp{(i\gamma/2)}, \\
        & \cos\gamma=\frac{\mu}{\sqrt{\mu^2+\eta^2}},
        \quad \sin\gamma=\frac{\eta}{\sqrt{\mu^2+\eta^2}}, \\
        & \eta=P_{\rm orbit}4\sigma \teq^3\kappa/c_p = \frac{P_{\rm orbit}}{\trad}.
    \end{split}
\end{equation*}
The TOA outgoing thermal flux perturbation is $F'(\tau=0,t)=(C+Z)\Delta F_v\exp(i \omega t)$, and the temperature
 perturbation at an arbitrary $\tau$ is
\begin{equation}
\begin{split}
    & T'(\tau,t) = \frac{\kappa\Delta F_v}{i\omega c_p}\exp(i\omega t)\times\\
    &[-kC\exp(-k\tau)-\beta Z\exp(-\beta\tau)+\beta\exp(-\beta\tau)].
    \label{eq.t'}
    \end{split}
\end{equation}

\begin{figure*}      
\epsscale{1.15}      
\centering
\plotone{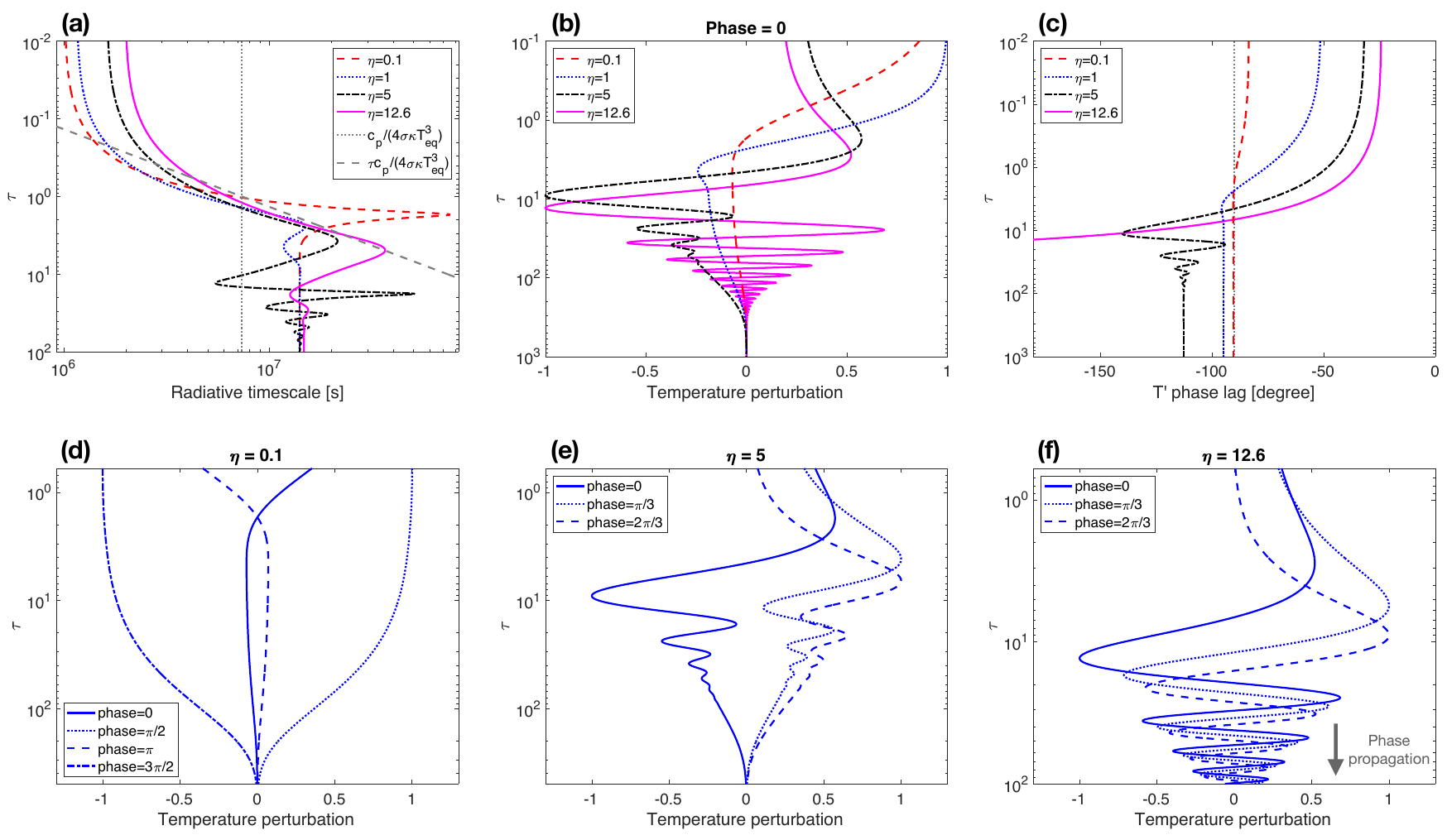}
\caption{Analytic calculations of semi-grey perturbations with different $\eta=P_{\rm orbit}/\trad$.  All cases assume an equilibrium temperature of $T_{\rm eq}=250$ K, a bond albedo $A=0.3$, a thermal opacity of $\kappa=5\times10^{-4}\;{\rm m^2kg^{-1}}$, a visible opacity of  $\kappa_v=2\times10^{-4}\;{\rm m^2kg^{-1}}$, $\mu_v=1$, and the specific heat of $c_p=13000 \;{\rm Jkg^{-1}K^{-1}}$. \emph{Panel (a):} radiative timescale calculated as $|T'|/|dT_{\rm th}/dt|$ where $dT_{\rm th}/dt = \frac{\kappa}{c_p}\frac{dF'}{d\tau}$ and $F'$ is given by Eq. (\ref{eq.solution}) as a function of optical depth $\tau$ for a few cases with different $\eta$. The grey dotted line remarks the value of $\frac{c_p}{4\sigma \kappa\teq^3}$ and the grey dashed line represents a linear relation $\frac{\tau c_p}{4\sigma \kappa\teq^3}$. \emph{Panel (b):}  temperature perturbations, that is the real part of $T'$, at phase 0 ($\omega t = 2n\pi$ where $n$ is an integer) as a function of $\tau$ with different $\eta$. These perturbations are greatly amplified and normalized so it is possible to visualize. The profiles with $\eta=0.1, 1$ and 5 are multiplied by $\exp{(0.39\tau)}$ and the profile with $\eta=12.6$ is multiplied by $\exp{(0.27\tau)}$. \emph{Panel (c):} phase lags of $T'$ as a function of $\tau$.  \emph{Panel (d), (e) and (f):} temperature perturbations as a function of $\tau$ at different phases for cases with different $\eta$. These perturbations are amplified and normalized as mentioned for panel (b).
}
\label{fig.appe}
\end{figure*}

{\bttt   For this particular problem we may naturally  define the radiative timescale as $|T'|/|dT_{\rm th}/dt|$, where $dT_{\rm th}/dt = \frac{\kappa}{c_p}\frac{dF'}{d\tau}$ is the heating rate caused solely by the perturbations of thermal radiation $F'$ given by Eq. (\ref{eq.solution}). Panel (a) in Figure \ref{fig.appe} shows the radiative timescales as a function of $\tau$ for a few different $\eta=P_{\rm orbit}/\trad$.  All cases assume parameters the same as those used in Section \ref{ch.semigrey}, including an equilibrium temperature of $T_{\rm eq}=250$ K, a bond albedo $A=0.3$, a thermal opacity of $\kappa=5\times10^{-4}\;{\rm m^2kg^{-1}}$, a visible opacity of  $\kappa_v=2\times10^{-4}\;{\rm m^2kg^{-1}}$, $\mu_v=1$, and the specific heat of $c_p=13000 \;{\rm Jkg^{-1}K^{-1}}$.  The radiative timescales are generally small at low $\tau$ and increase rapidly with $\tau$ toward the thermal photosphere. But  they stop increasing (some even decrease) and then converge to some values at greater depth. Interestingly, despite having the same atmospheric parameters, the profiles of the radiative timescale can quantitatively differ depending on the frequency of the seasonal irradiation. With this set of atmospheric parameters, the expression $\trad\sim\frac{p}{g}\frac{c_p}{4\sigma \teq^3}$, or $\trad\sim \frac{\tau c_p}{4\sigma \kappa\teq^3}$ (represented as the grey dashed line in panel (a)) if assuming a constant $\kappa$, that has been  commonly used in literature (e.g., \citealp{showman2002}) is a rather good approximation for those calculated by $|T'|/|dT_{\rm th}/dt|$ near the thermal photosphere. The value at $\tau=1$,  $\frac{c_p}{4\sigma \kappa\teq^3}$, which is plotted as the dotted grey line in panel (a), crosses the profiles calculated by $|T'|/|dT_{\rm th}/dt|$ in between $\tau=1$ to $1.4$ for different cases. These comparisons strongly support the use of $\trad\sim \frac{c_p}{4\sigma \kappa\teq^3}$ in Section \ref{ch.timescale}. }

 We discuss some interesting features of the deep perturbations by the seasonal irradiation. These perturbations are small and difficult to observe. In addition, temperature anomalies could be dominated by other sources including dynamics and diabatic effects from condensation/evaporation in the deep layers. Therefore, the following discussion is mostly for theoretical interest within our  framework.   Panel (b) in Figure \ref{fig.appe} shows the temperature perturbations, the real parts of $T'$, at phase 0 ($\omega t=2n\pi$ where $n$ is an integer) for the same set of cases. These perturbations are greatly amplified as well as normalized so that it is possible to visualize. The oscillations at the deep layers of cases with low and moderate $\eta$ exhibit a mode with a single long vertical wavelength. A higher $\eta=5$ shows a mixture of the long  mode and a short mode. The case with $\eta=12.6$ is dominated by the mode with short vertical wavelengths, showing many alternating perturbations at depth. Panel (c) shows the phase lags of $T'$ as a function of $\tau$ for  these cases. The phase lags of all cases are relatively large at low $\tau$  and decrease with increasing $\tau$. These phase lags all go across  $-90^{\circ}$ from low to high $\tau$ and this crossing occurs at higher $\tau$ for higher $\eta$. After the $-90^{\circ}$ crossing, the phase lag  with $\eta=0.1$ quickly converges to nearly $-90^{\circ}$ at high optical depth; the one with $\eta=1$ converges to a slightly smaller value at depth; the one with $\eta=5$ first shows  oscillations then converges to a value $<-100^{\circ}$ at depth; the one with $\eta=12.6$ continuously  decreases with increasing $\tau$.

{\bttt Time evolution of the deep layers  is depicted in panels (d), (e) and (f) in Figure \ref{fig.appe}, which show the temperature perturbations at different phases for cases with $\eta=0.1, 5$ and 12.6.  In the case with a low $\eta=0.1$, the long vertical mode appears to oscillate left and right. In the case with a high $\eta=12.6$, the short vertical mode appears to propagate downward with time.  The case with a medium $\eta=5$ shows a mixture of  the long mode that oscillates left and right and a short mode that propagates downward.  Note that the numerical inversion of the phase function generates phase lags within the range of ($-\pi, \pi$). For the case with $\eta=12.6$ (or other cases with $\eta\gtrsim 10$), the downward propagation of the short mode implies  that the phase lag will continuously decrease with increasing $\tau$ rather than oscillating in between ($-\pi, \pi$). In the phase lag for $\eta=12.6$ shown in panel (c), with increasing $\tau$, once the phase lag jumps to a value that is $2\pi$ greater than the adjacent value, it is subtracted by $2\pi$ to ensure continuity.  }

{\bttt Understanding of these behaviours can be obtained from limiting cases of the solution Eq. (\ref{eq.t'}). With a general $\eta/\mu$, at phase 0, temperature perturbation is  positive  at low optical depth and encounters a first transition to a negative value at layers with $\tau\gtrsim 1$ which also comes with a phase lag transition to $<-90^{\circ}$. The critical $\tau$ where this transition occurs is larger for case with   higher $\eta$.  This remarks the transition  from a somewhat direct response to irradiation to a thermally diffusive response at higher optical depth where the irradiation cannot reach and the oscillation is maintained by the diffusive thermal flux. The phase lag that is smaller than $-90^{\circ}$ cannot be obtained in the relaxation method discussed in Section \ref{ch.timescale}, which represents a qualitative difference between the relaxation and radiative transfer calculations.  Whether the deep oscillation is dominated by the long or short mode  then depends on properties of the solution at $\tau\gg 1$.  

When $\eta/\mu \ll 1$, $\cos{\gamma}\sim 1$, $\sin{\gamma}\sim \eta/\mu$, $k\sim (1+i\eta/(2\mu))/\mu$, $|Z|\ll 1$; because $|k|\sim 1/\mu>\beta$ for parameters used in Figure \ref{fig.appe}, $T'$ is dominated by terms of $\exp{(-\beta \tau)}$ at $\tau\gg1$: }
\begin{equation}
T'\sim \frac{\kappa\Delta F_v}{i\omega c_p}\beta(1-Z)\exp(-\beta\tau)\exp(i\omega t).
\label{eq.t'app}
\end{equation}
{\bttt Therefore, when $\eta/\mu\ll 1$, the real part of $T'$ does not change sign and $T'$ approaches a phase lag of nearly $-90^{\circ}$ at $\tau\gg 1$, corresponding to the long mode shown in Figure \ref{fig.appe}. }

{\bttt In the other limit of $\eta/\mu\gg 1$, $\gamma\sim\pi/2$, $k\sim(\eta\mu)^{-1/2}(1/\sqrt{2}+i/\sqrt{2})$. As $|k|<\beta$, at high optical depth, $T'$ is dominated by }
\begin{equation}
    T'\sim -\frac{\kappa\Delta F_v}{i\omega c_p}kC\exp(-k\tau)\exp(i\omega t),
\label{eq.t'app2}
\end{equation}
{\bttt in which the nonzero imaginary part of $k$ generates oscillations as a function of $\tau$, corresponding to the downward propagation of short vertical modes. The vertical wavelength is larger with larger $\eta$.  }

{\bttt For a moderate $\eta/\mu$, as long as $|k|>\beta$, the long mode given by Eq. (\ref{eq.t'app}) still holds at $\tau\gg1$ with a phase lag converging  to values $<-90^{\circ}$. But at medium $\tau$, the term with $\exp(-k\tau)$ can still contribute short modes in addition to the long mode, and the typical behavior is a mixture of a long and short modes as shown for the case with $\eta=5$ in Figure \ref{fig.appe}. With the set of parameters used here, the transition where $|k|<\beta$ occurs at $\eta\gtrsim 10$, explaining the transition from $\eta=5$ to $\eta=12.6$ in Figure \ref{fig.appe}.}

{\bttt The physical intuition is that in order to excite the short vertical modes,  the oscillation period of the irradiation has to exceed the radiative timescale slightly below the thermal photosphere which is typically a factor of several of $\frac{c_p}{4\sigma \kappa\teq^3}$ for parameters used for Figure \ref{fig.appe}. The significant vertical variation of radiative timescale near the thermal photosphere generates a large vertical phase lag variation over a short vertical distance. Only when the irradiation varies slower  than the local radiative timescale, the short vertical phase  variation  has time to propagate downward diffusively without being disrupted by the irradiation.}

{\bttt In cases with a higher $\beta=\kappa_v/(\kappa\mu_v)$ which can be achieved by  strong visible opacity, $|k|<\beta$ may be always satisfied and there will be short vertical modes at $\tau\gg1$ with arbitrary $\eta$. In such cases the calculated radiative timescales $|T'|/|dT_{\rm th}/dt|$ show overall smaller values for smaller $\eta$. }

\section{Analytic solution of non-grey perturbations}
\label{ch.appendix}

The radiative transfer equations with absorption only for the (picket fence) two characteristic thermal bands are:
\begin{equation}
\begin{split}
    \mu\frac{dI_1^{+}}{d\tau_1} &= I_1^{+} - (1-\chi)B(\tau_1), \\
    -\mu\frac{dI_1^{-}}{d\tau_1} &= I_1^{-} - (1-\chi)B(\tau_1), \\
    \mu\frac{dI_2^{+}}{d\tau_2} &= I_2^{+} - \chi B(\tau_2), \\
    -\mu\frac{dI_2^{-}}{d\tau_2} &= I_2^{-} - \chi B(\tau_2),
    \label{eq.nongreyrt1}
    \end{split}
\end{equation}
{\btt where $\chi$ is the fractional band width associated with $\kappa_2$ with respect to the total band width.} Similar to Appendix \ref{ch.appendix1}, we solve for the structure of perturbations around the radiative equilibrium.
Then we consider partitioning of irradiative energy into multiple channels in the visible band with different opacities $\kappa_{v,j}$ where $j=1,2, ...$. The perturbation of the irradiation in this case is treated as 
\begin{equation}
\begin{split}
  F'^-_v(t) & =\sum_j \Delta F_v \epsilon_j \exp{(i\omega t)} \exp{\left(-\frac{\tau_{v,j}}{\mu_v}\right)} \\
 & =  \Delta F_v  \exp{(i\omega t)} \sum_j \epsilon_j \exp(-\beta_j\tau_1), 
 \end{split}
 \label{eq.nongreyirr}
\end{equation}
where $\beta_j=\kappa_{v,j}/(\kappa_1\mu_v)$ and $\sum_j \epsilon_j=1$.

Writing the perturbations of the net thermal flux in band 1 and 2 as $\Tilde{F}'_1 = F'^+_1-F'^-_1$ and $\Tilde{F}'_2 = F'^+_2-F'^-_2$, together with equations (\ref{eq.dTdt})  and (\ref{eq.nongreyirr}), the perburbation form of equations (\ref{eq.nongreyrt1}) is organized as
\begin{equation}
\begin{split}
 \mu\frac{d^3\Tilde{F}'_1}{d\tau_1^2dt}= &\frac{1}{\mu}\frac{d\Tilde{F}'_1}{dt} - 8\pi\sigma T_{\rm RE}^3(1-\chi)\frac{\kappa_1}{c_p} \\
  & \left(\frac{d^2 \Tilde{F}'_1}{d\tau^2_1}+\frac{d^2\Tilde{F}'_2}{d\tau^2_1}-\Delta F_v \exp(i\omega t) \sum_j[\epsilon_j\beta_j^2\exp(-\beta_j\tau_1)]\right) \\
  \mu \mathcal{R}^2 \frac{d^3\Tilde{F}'_2}{d\tau_1^2dt}= & \frac{1}{\mu}\frac{d\Tilde{F}'_2}{dt} - 8\pi\sigma T_{\rm RE}^3\chi\mathcal{R}\frac{\kappa_1}{c_p} \\
  & \left(\frac{d^2 \Tilde{F}'_1}{d\tau^2_1}+\frac{d^2\Tilde{F}'_2}{d\tau^2_1}-\Delta F_v \exp(i\omega t) \sum_j[\epsilon_j\beta_j^2\exp(-\beta_j\tau_1)]\right),
\end{split}
\label{eq.nongreyrt2}
\end{equation}
Where $\mathcal{R}=\kappa_1/\kappa_2$. The time-dependent flux perturbations can be written as $\Tilde{F}'_1(\tau_1,t) = F'_1(\tau_1)\exp(i\omega t)$ and $\Tilde{F}'_2(\tau_1,t) = F'_2(\tau_1)\exp(i\omega t)$, and the above equation set becomes a set of coupled nonhomogeneous second-order differential equations with constant coefficients (again, we assume that $T_{\rm RE}$ is a constant over $\tau$ and $T_{\rm RE}=T_{\rm eq}$). We first seek general solutions to the homogeneous equation set and then obtain particular solutions for the full set. The complete solution is the combination of the general solutions and the particular solutions together with boundary conditions. The homogeneous part of Equations (\ref{eq.nongreyrt2}) can be written as
\begin{equation}
    \begin{split}
        \mathcal{A}_1\frac{d^2F'_1}{d\tau_1^2} = F'_1+\mathcal{A}_2\frac{d^2F'_2}{d\tau_1^2}\\
        \mathcal{B}_1\frac{d^2F'_2}{d\tau_1^2} = F'_1+\mathcal{B}_2\frac{d^2F'_1}{d\tau_1^2},
    \end{split}
    \label{eq.homogeneous}
\end{equation}
with constants
\begin{equation*}
    \begin{split}
        &\mathcal{A}_1\equiv\mu^2-i\eta\mu(1-\chi),\quad \mathcal{A}_2\equiv i\eta\mu(1-\chi),\\
        &\mathcal{B}_1\equiv \mu^2\mathcal{R}^2-i\eta\mu\chi\mathcal{R},\quad \mathcal{B}_2\equiv i\eta\mu\chi\mathcal{R},
    \end{split}
\end{equation*}
where $\eta=P4\sigma \teq^3\kappa_1/c_p = P_{\rm orbit}/\trad$. The general solution to the coupled equations (\ref{eq.homogeneous}) is  a linear combination of exponential functions on the form $F'_1=a\exp(k\tau_1)$ and $F'_2=b\exp(k\tau_1)$. By substituting these characteristic forms to equations (\ref{eq.homogeneous}), we obtain four values of $k$. Since perturbations should vanish when $\tau_1\rightarrow\infty$, only the set of $k$ with negative real parts is retained, which is
\begin{equation}
    k_{1,2}=-\left( \frac{-(\mathcal{A}_1+\mathcal{B}_1)\pm\sqrt{(\mathcal{A}_1-\mathcal{B}_1)^2+4\mathcal{A}_2\mathcal{B}_2}}{2(\mathcal{A}_2\mathcal{B}_2-\mathcal{A}_1\mathcal{B}_1)} \right)^{1/2}.
\end{equation}
The constants are
\begin{equation}
    b_{1,2} = a_{1,2}\frac{\mathcal{B}_2k_{1,2}^2}{\mathcal{B}_1k_{1,2}^2-1},
\end{equation}
with $a_{1,2}$ indeterminate.

Then we seek particular solutions to the full equation set (\ref{eq.nongreyrt2}) in the form of $F^{\ast}_1\exp(i\omega t)$ and $F^{\ast}_2\exp(i\omega t)$, with
\begin{equation}
\begin{split}
F^{\ast}_1=\Delta F_v\sum_j[\epsilon_j \mathcal{Z}_{1,j}\exp(-\beta_j\tau_1)] \\
F^{\ast}_2=\Delta F_v\sum_j(\epsilon_j \mathcal{Z}_{2,j}\exp(-\beta_j\tau_1)].
\end{split}
\label{eq.special}
\end{equation}
Substituting equations (\ref{eq.special}) to (\ref{eq.nongreyrt2}), we obtain coefficients as follows
\begin{equation}
        \mathcal{Z}_{1,j} = \frac{\mathcal{A}_2\beta_j^2(1-\mathcal{B}_1\beta_j^2-\mathcal{B}_2\beta_j^2)}{(\mathcal{A}_1\mathcal{B}_1-\mathcal{A}_2\mathcal{B}_2)\beta_j^4-(\mathcal{A}_1+\mathcal{B}_1)\beta_j^2+1} 
\end{equation}
        and
\begin{equation}
        \mathcal{Z}_{2,j}=\frac{\mathcal{B}_2\beta_j^2}{\mathcal{B}_1\beta_j^2-1}(\mathcal{Z}_{1,j}-1).
\end{equation}
The complete solution of the time-independent part of the net thermal flux is
\begin{equation}
\begin{split}
F'_1=a_1\exp(k_1\tau_1)+a_2\exp(k_2\tau_1)+F^{\ast}_1,\\
F'_2=b_1\exp(k_1\tau_1)+b_2\exp(k_2\tau_1)+F^{\ast}_2,
\end{split}
\label{eq.solution_nongrey}
\end{equation}
in which $a_{1,2}$ are yet to be determined. 
We make use the upper boundary conditions $F'^-_1=0$ and $F'^-_2=0$ to determine $a_{1,2}$, yielding
\begin{equation}
\begin{split}
& a_1 = \frac{ \Delta F_v\sum_j\left(\epsilon_j\mathcal{Z}_{2,j}\frac{\beta_j\mathcal{R}\mu+1}{\beta_j\mathcal{R}\mu}-\epsilon_j \mathcal{Z}_{1,j}\frac{\beta_j\mu+1}{\beta_j\mu}\frac{\mathcal{B}_2k^2_2}{\mathcal{B}_1k^2_2-1}\frac{1-k_2\mu\mathcal{R}}{\mathcal{R}(1-k_2\mu)}   \right) }
{\frac{\mathcal{B}_2k_1}{\mathcal{B}_1k^2_1-1}\frac{1-k_1\mu\mathcal{R}}{\mu\mathcal{R}}-\frac{\mathcal{B}_2k^2_2}{\mathcal{B}_1k^2_2-1} \frac{1-k_2\mu\mathcal{R}}{\mathcal{R}(1-k_2\mu)}\frac{1-k_1\mu}{k_1\mu}  }\\
& a_2 = \frac{ \Delta F_v\sum_j\left(\epsilon_j\mathcal{Z}_{2,j}\frac{\beta_j\mathcal{R}\mu+1}{\beta_j\mathcal{R}\mu}-\epsilon_j \mathcal{Z}_{1,j}\frac{\beta_j\mu+1}{\beta_j\mu}\frac{\mathcal{B}_2k^2_1}{\mathcal{B}_1k^2_1-1}\frac{1-k_1\mu\mathcal{R}}{\mathcal{R}(1-k_1\mu)}   \right) }
{\frac{\mathcal{B}_2k_2}{\mathcal{B}_1k^2_2-1}\frac{1-k_2\mu\mathcal{R}}{\mu\mathcal{R}}-\frac{\mathcal{B}_2k^2_1}{\mathcal{B}_1k^2_1-1} \frac{1-k_1\mu\mathcal{R}}{\mathcal{R}(1-k_1\mu)}\frac{1-k_2\mu}{k_2\mu}  }.
\end{split}
\label{eq.a12}
\end{equation}
Finally, the desired solution for thermal fluxes is that in Eqs. (\ref{eq.solution_nongrey}) times $\exp(i\omega t)$.
Temperature perturbation as a function of time and optical depth $\tau_1$ is derived using 
\begin{equation}
    \frac{dT}{dt} = \frac{g}{c_p}\frac{dF_{\rm tot}}{dp} = \frac{\kappa_1}{c_p}\frac{dF_{\rm tot}}{d\tau_1},
\end{equation}
which is given by
\begin{equation}
    \begin{split}
        T(\tau_1,t) &= \frac{\kappa_1}{i\omega c_p}\frac{\partial F'_{\rm tot}}{\partial\tau_1}  
        =\frac{\kappa_1}{i\omega c_p}\exp(i\omega t)\times \\
        & \{ a_1k_1\exp(k_1\tau_1)+a_2k_2\exp(k_2\tau_1) - \Delta F_v\sum_j[\epsilon_j\mathcal{Z}_{1,j}\beta_j\exp(-\beta_j\tau_1)] \\
        &+ b_1k_1\exp(k_1\tau_1)+b_2k_2\exp(k_2\tau_1) - \Delta F_v\sum_j[\epsilon_j\mathcal{Z}_{2,j}\beta_j\exp(-\beta_j\tau_1)] \\
        & +\Delta F_v\sum_j[\epsilon_j\beta_j\exp(-\beta_j\tau_1)] \}.
    \end{split}
\end{equation}
Likewise, these solutions can be expressed  as a combination of amplitudes and phase lags.

{\bttt The perturbations in the deep layers in the non-grey situation should have qualitatively similar properties as those in the semi-grey cases shown in Appendix \ref{ch.appendix1}. This is because at high optical depth both cases collapse into  the diffusive limit which may be adaptively described by the semi-grey framework.  }

\end{document}